\documentclass{article}

\usepackage{arxiv}

\usepackage[utf8]{inputenc} % allow utf-8 input
\usepackage[T1]{fontenc}    % use 8-bit T1 fonts
\usepackage{hyperref}       % hyperlinks
\usepackage{url}            % simple URL typesetting
\usepackage{booktabs}       % professional-quality tables
\usepackage{amsfonts}       % blackboard math symbols
\usepackage{nicefrac}       % compact symbols for 1/2, etc.
\usepackage{microtype}      % microtypography
\usepackage{lipsum}
\usepackage{graphicx}
\usepackage{enumitem}
\usepackage{multirow}
\usepackage[export]{adjustbox}
\usepackage{natbib}
\usepackage{xcolor} % Yuvraj added this for my annotations in color. 
\usepackage{array}
\newcolumntype{L}[1]{>{\raggedright\let\newline\\\arraybackslash\hspace{0pt}}m{#1}}
\newcolumntype{R}[1]{>{\raggedleft\let\newline\\\arraybackslash\hspace{0pt}}m{#1}}
\newcolumntype{C}[1]{>{\centering\let\newline\\\arraybackslash\hspace{0pt}}m{#1}}

\title{Decentralized is not risk-free: Understanding public perceptions of privacy-utility trade-offs in COVID-19 contact-tracing apps}

\author{
  Tianshi Li \\
  Carnegie Mellon University\\
  Pittsburgh, PA 15213 \\
  \texttt{tianshil@cs.cmu.edu} \\
   \And
 Jackie (Junrui) Yang \\
  Stanford University\\
  Stanford, CA 94305 \\
  \texttt{jackiey@cs.stanford.edu} \\
   \And
 Cori Faklaris \\
  Carnegie Mellon University\\
  Pittsburgh, PA 15213 \\
  \texttt{cfaklari@cs.cmu.edu} \\
   \And
 Jennifer King \\
  Stanford University\\
  Stanford, CA 94305 \\
  \texttt{jenking@law.stanford.edu } \\
  \And
 Yuvraj Agarwal \\
  Carnegie Mellon University\\
  Pittsburgh, PA 15213 \\
  \texttt{yuvraj@cs.cmu.edu} \\
  \And
 Laura Dabbish \\
  Carnegie Mellon University\\
  Pittsburgh, PA 15213 \\
  \texttt{dabbish@cs.cmu.edu} \\
  \And
 Jason I. Hong \\
  Carnegie Mellon University\\
  Pittsburgh, PA 15213 \\
  \texttt{jasonh@cs.cmu.edu} \\
}

\begin{document}
\maketitle

\begin{abstract}
Contact-tracing apps have potential benefits in helping health authorities to act swiftly to halt the spread of COVID-19.
However, their effectiveness is heavily dependent on their installation rate, which may be influenced by people's perceptions of the utility of these apps and any potential privacy risks due to the collection and releasing of sensitive user data (e.g., user identity and location).
In this paper, we present a survey study that examined people's willingness to install six different contact-tracing apps after informing them of the risks and benefits of each design option (with a U.S.-only sample on Amazon Mechanical Turk, $N=208$).
The six app designs covered two major design dimensions (centralized vs decentralized, basic contact tracing vs. also providing hotspot information), grounded in our analysis of existing contact-tracing app proposals.

Contrary to assumptions of some prior work, we found that the majority of people in our sample preferred to install apps that use a centralized server for contact tracing, as they are more willing to allow a \textit{centralized authority} to access the identity of \textit{app users} rather than allowing \textit{tech-savvy users} to infer the identity of \textit{diagnosed users}.
We also found that the majority of our sample preferred to install apps that share diagnosed users' recent locations in public places to show hotspots of infection.
Our results suggest that apps using a centralized architecture with strong security protection to do basic contact tracing and providing users with other useful information such as hotspots of infection in public places may achieve a high adoption rate in the U.S.
We also offer some recommendations on how to communicate the risks and benefits of contact tracing apps with the general public.
\end{abstract}

% keywords can be removed

\section{Introduction}

Contact tracing is an important approach for dealing with contagious diseases such as COVID-19~\cite{10.1371/journal.pone.0000012}.
Contact tracing involves tracing and monitoring contacts of infected people for identifying and supporting exposed individuals to be quarantined on time.
However, manual contact tracing requires a large investment in human resources and does not always achieve precise tracing results~\cite{SomeStat73:online}.

Digital contact-tracing apps~\cite{chan2020pact} can potentially alleviate the burden on human contact tracers and improve tracing precision.
Given the high adoption rate of smartphones (currently over 80\% in the U.S. in 2020~\cite{Smartpho80:online}) and their rich sensing capabilities (e.g., BLE-based proximity sensing and location-based tracking), these apps can automate the laborious work of tracing back users' contact history to help find who an infected person has been in contact with recently (e.g., in the last 14 days.)
These apps can provide notice to users of their potential exposure to the virus and guide them to take tests and to self-quarantine.
They can also gather information to help epidemiologists monitor the spread of the disease, discover disease hotspots, and contact exposed individuals. 
However, their effectiveness is highly dependent on the adoption rate, which has been demonstrated to be challenging due to people's concerns about issues such as privacy~\cite{PrimeMin96:online, Bluetoot90:online, 2005.04343}.
Ferretti et al.~\cite{10.1126/science.abb6936,Digitalc69:online} suggested that if 60\% of the population installed the app, the estimated number of coronavirus cases would go down.

To achieve accurate contact tracing and provide timely exposure notice, apps need to collect sensitive user information such as one's location history and contact information.
With more information collected, the app can provide more functionality.
For example, the app can release the aggregated whereabouts of infected users to help the public gauge the risk of going to certain areas.
Also, the app may require users to provide their real identity information to better integrate the tool into part of the normal workflow of contact tracers.
However, the more information that is disclosed to the app, the greater the privacy risks that users of the app can be exposed to.
This involves a classic problem in privacy, namely how people manage tradeoffs between privacy and utility.
In this problem, it is particularly essential to gain a better understanding into public perceptions of these trade-offs, because they may have a significant impact on the adoption rate of these apps.

Although there are many attempts to create COVID-19 contact-tracing apps in a privacy-preserving way \cite{GitHubDP97:online, chan2020pact}, it is less clear whether sufficient people are willing to install the apps, and what design choices can lead to higher adoption rates.
Some recent studies have started looking into this problem by projecting the installation rate of contact-tracing apps in the U.S. in general \cite{Washingt23:online, American37:online} or examining how the willingness to adopt COVID-19 apps varies with accuracy levels, public health benefits, and who the data may be leaked to \cite{2005.04343}.
However, none of them has connected these abstract factors to concrete app-design choices to provide guidelines on designing and deploying COVID-19 apps that can achieve optimal adoption rates.
There is a large design space for contact-tracing apps, each with different functionality as well as implications for privacy.
Evaluating these differences can help us better understand how people view the privacy-utility trade-offs, and lead to better designs that more people are likely to adopt.

In this paper, we present a survey study with a U.S.-only sample that aims to bridge this gap by placing the choices in the hands of the general public:
\emph{If provided with a set of options and informed of their privacy risks and utility benefits, what design of contact-tracing app would make people agree or disagree to install the app? }
To unpack the complex trade-offs in this problem, we identified two important design dimensions that can affect the privacy-utility trade-off based on existing contact-tracing app designs (Table \ref{tab:design_dimensions}).

\textbf{The first dimension} characterizes whether the app uses a proximity-based~\footnote{We have not considered location-based content contact tracing for the sake of simplicity.} centralized or decentralized architecture to achieve basic contact tracing.
In both architectures, users' phones can generate many ephemeral identifiers based on user identifiers, broadcast them using a channel (BLE or ultrasound), and record the broadcasted identifiers they have received.
In a centralized architecture, contact tracing is done on centralized servers that ask \emph{infected users} to upload all the broadcasted identifiers they have received.
Then they can decode them to the identities of exposed users, which allows health authorities to notify these users.
In this setup, users will only know one bit of information: whether they were recently exposed to an infected user or not.
Note that we only discuss centralized architecture that requires user's identity in this paper (e.g., phone number in BlueTrace~\cite{bay2020bluetrace}), since the other alternatives (such as ROBERT~\cite{document21:online, document64:online}) provide few additional utility benefits while introducing more privacy risks (see Section \ref{sec:why_centralized_identity}).
Therefore, for centralized contact-tracing apps, the identity information of app users and their status (infected, exposed, or other) may be accessible by health workers, app developers, and state/federal-level health authorities (depending on at what level the app is deployed). 

In a decentralized architecture, central servers collect diagnosed users' identifiers (e.g., Diagnosis Keys in Privacy-preserving contact tracing \cite{Appleand45:online}), and push them to every registered user.
Contact tracing is then done in a decentralized way on each user's phone by comparing the infected users' identifiers to the broadcasts recorded locally.
The identity information of app users will remain private from the central authorities, while tech-savvy users may have the ability to infer the identities of the diagnosed users that they have been in close proximity to (see Section~\ref{sec:tech_savvy_infer_identity}).
% In contrast, with decentralized architectures, users will receive a full list of pseudonymized identifiers of infected users to complete contact tracing, which may be re-identifiable.
% Because when users are diagnosed with the disease and have given consent, a subset of their pseudonymized identifiers (e.g., Diagnosis Keys in Privacy-preserving contact tracing \cite{Appleand45:online}) will be uploaded to the central server, and pushed to other users' devices to help them determine whether they have recently received proximity identifiers derived from these keys.

\begin{table}[h]
    \centering
    \caption{Our survey examines two essential design dimensions of contact-tracing apps that involve privacy-utility tradeoffs.
    Design dimension 1 (two choices): Centralized architecture vs. Decentralized architecture.
    Design dimension 2 (three choices): Do not collection locations vs. Collect locations (public places vs. all places) to provide infection hotspots to the public.}
    \begingroup
    \setlength{\tabcolsep}{7pt} % Default value: 6pt
    \renewcommand{\arraystretch}{2} % Default value: 1
    \begin{tabular}{p{0.1\linewidth}p{0.2\linewidth}p{0.22\linewidth}p{0.22\linewidth} p{0.15\linewidth}}
    \toprule
        \small{\textbf{Dimension}} & \small{\textbf{Design choice}} & \small{\textbf{Utility benefits}} & \small{\textbf{Privacy risks}} & \small{\textbf{Examples\protect\footnotemark}} \\
        \hline
        Centralized 
        
        vs.
        
        Decentralized & Centralized architecture & Health workers can contact exposed users and guide them to take tests and self-quarantine. & \emph{Health workers, health authorities} and \emph{app developers} may know who installed the app, who are \emph{infected}, and who have been \emph{exposed} to infected users. & TraceTogether
        
        (Singapore)~\cite{bay2020bluetrace},
        
        COVIDSafe
        
        (Australia)~\cite{COVIDSaf97:online},
        BlueTrace~\cite{bay2020bluetrace}
        \\
        & Decentralized architecture & The contact-tracing app can inform exposed users and provide guidance on how to take tests and self-quarantine. & \emph{Tech-savvy users} may be able to infer the identities of some \emph{infected} users they have been in contact with by logging additional location information \cite{de2013unique} or opening multiple accounts \cite{document64:online}. & DP-3T~\cite{GitHubDP97:online}
        
        (East) PACT~\cite{WaybackM97:online}
        
        (West) PACT~\cite{2004.03544}

        Germany app~\cite{GermanyW23:online}\\
        \midrule
        Location
        
        collection & Collect location history of infected users & The general public can know the whereabouts of infected users in \emph{all places}. & Everyone can view the aggregated location history of \emph{infected} users. & South Korea
        
        system~\cite{SouthKor69:online}\\
        & Collect location history of infected users in \emph{public areas} only (e.g., parks, restaurants) & The general public can know the whereabouts of infected users in \emph{public places}. & Everyone can view the aggregated location history in public areas of \emph{infected} users. &
        Private Kit~\cite{2003.08567} \\
        & Do not collect location & No additional utility benefits. & No additional privacy risks. & TraceTogether
        
        (Singapore)~\cite{bay2020bluetrace},
        
        COVIDSafe
        
        (Australia)~\cite{COVIDSaf97:online}\\
    \bottomrule
    \end{tabular}
    \endgroup
    \label{tab:design_dimensions}
\end{table}
\footnotetext{These examples have similar design choices but may not be exact match. They are listed to demonstrate that real-world apps sit in different positions across these spectra.}

In addition to basic contact tracing, some countries (e.g., mainland China, South Korea~\cite{Coronavi22:online}) also make the locations of infected users public so that people could stay away from disease hotspots and report to the contact tracers if they have been in these areas recently.
Therefore, our \textbf{second design dimension} characterizes whether and to what extent the app collects and shares user location to provide information about hotspots.
Our survey presents three representative design choices:
1) collect users' locations in \emph{all places} and release aggregated and coarse-grained location history of infected users;
2) collect users' locations when they are in \emph{public places} and release aggregated and coarse-grained location history of infected users;
3) do not collect users' locations and do not provide information about infection hotspots.

These two dimensions are independent of each other, which results in six designs of contact-tracing apps.
We designed our survey around these six options, probing people's attitudes about the apps' trade-offs.
We conducted the survey study on Amazon Mechanical Turk and collected 244 responses from unique MTurk workers (208 valid responses) from April 27, 2020 to May 7, 2020.
The HIT was restricted to only be visible to MTurk workers in the United States.
Our results yielded a number of findings on how people view privacy-utility trade-offs in these apps that may help design better COVID-19 apps that more people are willing to adopt, including:

\begin{itemize}
    \item While there were mixed attitudes towards different app design choices, overall, the centralized design option and the public-only location sharing option showed a significant positive effect on installation preferences (Section \ref{sec:overall_preference_results}).
    \item People seem to be more concerned about the risk of decentralized apps that may allow tech-savvy users to identify infected users, as compared to centralized apps that allow health workers and state-/federal-level health authorities to know user identities, which is contrary to what was suggested in prior work that the decentralized solutions have the best protection of users' privacy~\cite{GermanyW23:online,JointSta90:online}. (Section \ref{sec:privacy_related_factors} and \ref{sec:qualitative_codes})
    \item Centralized designs had a higher positive effect on installation preferences in blue states than red states (the division of red/blue states was based on the 2016 United States presidential election results). (Section \ref{sec:overall_preference_results})
    \item We found that about 25\% of our participants had particularly strong feelings about privacy and were unlikely to install any contact-tracing app regardless of any built-in privacy protections. These people seemed to prefer decentralized designs over centralized designs. This finding aligns with a study by Kaptchuk et al.~\cite{2005.04343} that found 27\% of people do not want to install COVID contact-tracing apps even if they are perfectly private (Section \ref{sec:clustering_analysis}).
    \item The most popular app design (centralized, release infection hotspots in public places) had around 55\% of our participants willing to install (Section \ref{sec:overall_preference_results}).
    This is similar to the results (around 50\% smartphone users agree to install) of the Washington Post and the Ipsos polls~\cite{Washingt23:online, American37:online}.
\end{itemize}

Based on these findings, we derived the following suggestions on designing COVID-19 apps that respect users' privacy and have more people willing to install (may only apply to situations in the U.S.):
\begin{itemize}
    \item \textbf{App Design: Centralized vs Decentralized:}
    Based on people's preferences of privacy-utility trade-offs, if a single COVID-19 contact-tracing app is going to be deployed in the U.S. at a national level, a proximity-based centralized architecture may be a better design option than proximity-based decentralized architecture.
    However, if different apps are going to be deployed for different states, then the varying preferences in states with different partisan leanings should be taken into account.
    If centralized solutions are adopted, the app should verify every user's identity to prevent malicious users from identifying infected users by signing up multiple accounts.
    Also it will be especially important to apply strong security protections to prevent data breaches, and make sure the collected data is only used for the purposes specified to users when requesting consent to access the data. 
    \item \textbf{App Design: Location Sharing:}
    In addition to supporting basic contact-tracing functionality, providing users with more useful information may nudge more people to install the app.
    For example, releasing the aggregated whereabouts of infected users has been suggested as a useful feature with acceptable privacy risks and considerable utility benefits.
    Due to the different levels of acceptance to sharing location history, as suggested by our clustering analysis results in Section \ref{sec:clustering_analysis}, the app should request the location collection in a progressive way, and provide users with sufficient control regarding to what extent their location could be released to the public (e.g., sharing all location, sharing location in public areas only, not sharing location, etc.) 
    \item \textbf{Design of Privacy Notices:}
    These apps should be transparent about the risks of disclosing personal information, both to governments (for centralized designs) and to tech-savvy users (for decentralized designs).
    Decentralized solutions should not be posited as a risk-free solution, since people seem to have more problem with tech-savvy users identifying them (as in decentralized designs) than health workers and health authorities doing so (as in centralized designs).
    % cite news sources: https://www.npr.org/sections/coronavirus-live-updates/2020/04/27/846046185/germany-backs-away-from-compiling-coronavirus-contacts-in-a-central-database
    % https://www.coindesk.com/european-contact-tracing-consortium-faces-wave-of-defections-over-centralization-concerns
\end{itemize}

% Talk about our assumption of the scenarios and what the app guarantees: 
% 1) the contact tracing process is done automatically with technology, instead of self-reporting location history or scanning QR code at certain location
% 2) each contact-tracing app is developed so that it collects the minimum data to support its functionality.

% 1) We assume the app is deployed at state or federal level. 2) We do not discuss options that allow infected users to redact their traces (e.g., DP-3T)

\section{Related work}

In this section, we 
%provide background from epidemiological studies on contact tracing, 
list the contact-tracing apps that we reviewed to derive our exemplar designs, and compare our study with prior studies about COVID-19 tracing apps.

% \jason{I think cut section 2.1, since it's too short to be meaningful. Perhaps move elements to the intro, essentially saying that many epidemiologists have strongly argued for digital contact tracing as one way of helping address COVID.}

% \subsection{Epidemiological studies on contact tracing}

% %Talk about the increased emphasis on this aspect from the society and health authorities, and the inadequacy of contact tracers in the current situation. Digital contact tracing technologies can potentially alleviate their burden.

% %Theoretical simulation regarding COVID-19 contact-tracing apps

% Contact tracing is generally a useful method for controlling emerging epidemics~\cite{10.1371/journal.pone.0000012}.
% For COVID-19 specifically, Hellewell et al.~\cite{10.1016/s2214-109x(20)30074-7} developed a transmission model which showed that if the $R_0$\footnotemark for COVID-19 is $2.5$, 70\% of the contacts had to be traced to effectively curb the virus.
% \footnotetext{$R_0$ is the basic reproduction number. $R_0$ is the expected number of people who will get infected generated by a single infected person in the population.}
% Ferretti et al.~\cite{10.1126/science.abb6936} investigated in the possibility of using digital contact-tracing apps.
% Their simulation showed that if about 60\% of the population use the contact-tracing apps, there will be a reduction in the number of coronavirus cases and deaths. 
% This proves the importance of contact-tracing apps as a measure to control the virus and the installation base of contact-tracing apps as a limiting factor of this measure.

\subsection{COVID-19 Apps}
%Here discuss major apps that are already built and deployed, as well as impactful research proposals.
%
%threat modeling

%\jason{I changed the subsection title to contact-tracing apps. Is this still correct, do  all the apps listed below do that?}

Several contact-tracing apps have been developed, each with different features and different personal data requirements.

Many governments around the world have released COVID-19 apps.
They are mostly focused on location history collection and BLE-based contact tracing with a centralized server. One example is
Health code~\cite{Chinaisf93:online}, which is used in part of mainland China. This app asks requires users to self-report symptoms, and also asks for their home and work addresses so as to warn about potential exposures. However, details of their algorithm have not been released.
TraceTogether and its BLE-based protocol BlueTrace~\cite{bay2020bluetrace}, made by the Singapore government, supports contact tracing by collecting IDs of exposed users and mapping the IDs to their phone numbers on a centralized server.
PEPP-PT~\cite{peppptdo62:online}, a joint effort between France, Germany, and Italy, proposed a centralized design using BLE-based proximity sensing.
South Korea's contact tracing solution~\cite{SouthKor69:online} uses the location history of infected patients as collected from cell towers to inform people that have potentially been in high-risk areas.
India's contact-tracing app CoWin-20~\cite{Indiaisb92:online} requires a user's position and BLE-based proximity data, but the details of their algorithm has not yet been released.
Germany tried to implement a centralized BLE-based proximity contact tracing at the beginning, but then switched to a decentralized solution~\cite{GermanyW23:online}.

On the academic side, many researchers~\cite{JointSta90:online} have argued that a decentralized infrastructure for contact tracing is better for end-user privacy.
Following this idea, Troncoso et al. proposed DP-3T~\cite{GitHubDP97:online}, Rivest et al. proposed PACT~\cite{2004.03544}, Gebhard et al.~\cite{TCNCoali46:online} proposed TCN-protocol, and Arx et al. proposed COVID Watch~\cite{HowItWor12:online}, all of which use a decentralized infrastructure and BLE-based proximity sensing to facilitate contact tracing.
Along these same lines, Apple and Google~\cite{Appleand45:online} have released the ``privacy-preserving contact tracing'' API, which can support building decentralized contact-tracing apps.
Loh et al. have proposed NOVID~\cite{NOVID17:online}, which uses ultrasound along with Bluetooth to improve the accuracy of physical proximity measurements.
Researchers also tried to make location collection more privacy-friendly.
Raskar et al. proposed Private Kit~\cite{2003.08567}, which is a location-based contact tracing solution that supports redaction of location traces to preserve privacy.
Prior work has also built apps that can collect a user's self-reported symptoms.
Spector et al. released COVID Symptom Study~\cite{Aboutthi42:online}, which also can be used to identify ``hotspots'' of infections from reported symptoms.
CoEpi~\cite{AboutCoE88:online} combined self-reported symptom with BLE-based proximity tracing solution so that it can warn the exposed user even before the official diagnosis.

We observed that most of the COVID-19 apps released are focused on location/proximity-based contact tracing~\cite{bay2020bluetrace,peppptdo62:online,Indiaisb92:online,GermanyW23:online,GitHubDP97:online,2004.03544,TCNCoali46:online,HowItWor12:online,Appleand45:online,NOVID17:online,2003.08567,AboutCoE88:online},
location-based hotspot reporting~\cite{SouthKor69:online,Indiaisb92:online,2003.08567},
and self-reported symptom tracking~\cite{Chinaisf93:online,Aboutthi42:online,AboutCoE88:online}.
In this paper we focused on apps that support proximity-based contact tracing and location-based hotspot reporting.
These contact-tracing apps can generally be divided by whether the contact tracing is done on a \textit{centralized} server~\cite{bay2020bluetrace,peppptdo62:online} or done on every user's phone in a \textit{decentralized} way~\cite{GitHubDP97:online,2004.03544,TCNCoali46:online,HowItWor12:online,Appleand45:online,NOVID17:online,AboutCoE88:online}.
Our survey investigates app designs for both types of apps, with the exception that we only consider centralized contact-tracing apps that require user's identity when signing-up.
Our rationale is detailed in Section \ref{sec:why_centralized_identity}.

\label{sec:tech_savvy_infer_identity}
Interestingly, we found that many papers tend to favor decentralized solutions because they share less sensitive data with central authorities.
However, the risks of decentralized solutions are not always discussed.
Since pseudonymized identifiers of infected users will be shared with all users, tech-savvy users potentially have the capability to re-identify  infected users if additional location data has been logged when the pseudonymized identifiers were received \cite{GitHubDP97:online, de2013unique}.
In fact, our survey results suggest that people are concerned about the risk of tech-savvy users knowing the identities of infected users, with many of our participants considering it less acceptable than health authorities knowing the same information.

\subsection{Studies about COVID-19 contact-tracing apps}
%Analyzing security threats, risks and benefits
%Surveys: WaPo \& UMD, Oxford, MSR

There have been some early studies of people's perceptions of COVID-19 contact-tracing apps.

A April 2020 poll conducted by the Washington Post and the University of Maryland~\cite{Washingt23:online} showed that 50\% of smartphone users will ``definitely'', or `probably'', use a contact-tracing app.
Another poll in May 2020 by Ipsos~\cite{American37:online} also suggested that 51\% Americans would join a CDC-sponsored cell phone-based contact tracing system.
Milsom et al.~\cite{Milsom_Abeler_Altmann_Toussaert_Zillessen_Blasone_2020} have conducted a survey about the general acceptability of app-based contact tracing among individuals residing in the UK, the U.S., France, Germany, and Italy.
Their results showed that the proportion of people in favor of installing a contact-tracing app ranged from 67.5\% to 85.6\%.
All three surveys focused on general preference on installing a contact-tracing app, but did not distinguish between different app designs.
Moreover, they did not mention the privacy risks of those apps, and therefore may not capture users' opinions when privacy issues are fully taken into account.

Kaptchuk et al.~\cite{2005.04343} investigated how different benefits, accuracy-levels, and privacy decisions affect user's installation rate.
They showed that 75\% to 80\% of people in the U.S. may install a perfectly private and accurate app.
The difference between Kaptchuk et al. and our work is that their studies are not rooted in concrete app design options, and was not able to provide suggestions on how to handle the trade-offs when perfect privacy and utility cannot be achieved at the same time.
In contrast, our survey directly probed users' preferences using six representative app designs and resulted in app design suggestions for achieving optimal adoption rates.

%\subsection{Perceptions of privacy in different populations}

\section{Methodology}
In this section, we present how we chose our six app design options, the design of the survey itself, and how we conducted the survey on MTurk.

\subsection{Making the Six Representative App Designs}
\label{sec:design_process_of_six_app_options}

Table \ref{tab:six_app_design_options} summarizes the six app design options covered in our survey.
These options are driven by the two design dimensions presented in Table \ref{tab:design_dimensions} (see the ``Design choice'' column.) 

Since we are targeting the general public, an important requirement of our survey design is to describe how these apps differ in terms of utility and privacy risks without being overly technical (see Figure \ref{fig:pairwise_comparison_viz}.)
Our approach was to first present the functionality that can be supported by the corresponding design choice, and then to present the privacy risks assuming that the minimum amount of data is collected for that case. We assumed no data breaches due to unauthorized access (e.g., central server being hacked).

\label{sec:six-stakeholders}
We referred to several existing analyses to determine the data practices of those app designs~\cite{GitHubDP97:online}, and mentioned the most essential data types and stakeholder types to present comprehensive and intelligible app descriptions to our participants.
We focused on how two basic types of data -- location and users' precise identities -- and whether that data might be directly or indirectly shared with six stakeholders: health workers, app developers, state-level/federal-level health authorities, tech-savvy users, and the general public.

\begin{table}[]
    \centering
    \caption{The six app design options abstracted based on design choices from the two design dimensions presented in Table \ref{tab:design_dimensions}. App 1-2 and 4-5 only collect coarse-grained location history (e.g., \char`~1000m/3000ft accuracy), and release infected users' location history to the public. Note that not all six app designs can match exactly with a real world app. Some may follow a similar design (e.g., the Germany contact-tracing app that used a centralized design), but had certain flaws in their implementation which could cause more data to be leaked \cite{GitHubDP97:online}.}
    \begin{tabular}{p{0.03\linewidth} p{0.12\linewidth} p{0.3\linewidth} p{0.45\linewidth}}
    \toprule
        \small{\textbf{App}} & \small{\textbf{Design choices}} & \small{\textbf{Utility}} & \small{\textbf{Minimum required data}}  \\
        \midrule
        1 & Centralized + all loc. & \vspace{-1\baselineskip}\begin{itemize}[leftmargin=*]
            \item Health workers inform exposure
            \item Show hotspots in \emph{all places} to the public
        \end{itemize}&
          \vspace{-1\baselineskip}\begin{itemize}[leftmargin=*]
            \item Infected users' coarse-grained location history $\rightarrow$ 
            
            \{health workers, app developers, state/federal health authorities, the public\}
            \item All users' identities $\rightarrow$
            
            \{health workers, app developers, state/federal health authorities\}
        \end{itemize}\\
        2 & Centralized + public loc. & \vspace{-1\baselineskip}\begin{itemize}[leftmargin=*]
            \item Health workers inform exposure
            \item Show hotspots in \emph{public} places to the public
        \end{itemize} & \vspace{-1\baselineskip}\begin{itemize}[leftmargin=*]
            \item Infected users' coarse-grained location history in public areas $\rightarrow$ 
            
            \{health workers, app developers, state/federal health authorities, the public\}
            \item All users' identities $\rightarrow$
            
            \{health workers, app developers, state/federal health authorities\}
        \end{itemize}\\
        3 & Centralized + no loc. & \vspace{-1\baselineskip}\begin{itemize}[leftmargin=*]
            \item Health workers inform exposure
            \item No hotspot information available to the public
        \end{itemize} & \vspace{-1\baselineskip}\begin{itemize}[leftmargin=*]
            \item All users' identities $\rightarrow$ 
            
            \{health workers, app developers, state/federal health authorities\}
            \end{itemize} \\
        4 & Decentralized + all loc. & \vspace{-1\baselineskip}\begin{itemize}[leftmargin=*]
            \item App informs exposure
            \item Show hotspots in \emph{all places} to the public
        \end{itemize} & \vspace{-1\baselineskip}\begin{itemize}[leftmargin=*]
            \item Infected users' coarse-grained location history $\rightarrow$ 
            
            \{health workers, app developers, state/federal health authorities, the public\}
            \item Infected users' identities $\rightarrow$
            
            \{tech-savvy users\}
        \end{itemize} \\
        5 & Decentralized + public loc. & \vspace{-1\baselineskip}\begin{itemize}[leftmargin=*]
            \item App informs exposure
            \item Show hotspots in \emph{public places} to the public
        \end{itemize} & \vspace{-1\baselineskip}\begin{itemize}[leftmargin=*]
            \item Infected users' coarse-grained location history in public areas $\rightarrow$
            
            \{health workers, app developers, state/federal health authorities, the public\}
            \item Infected users' identities $\rightarrow$
            
            \{tech-savvy users\}
        \end{itemize} \\
        6 & Decentralized + no loc. &  \vspace{-1\baselineskip}\begin{itemize}[leftmargin=*]
            \item App informs exposure
            \item No hotspot information available to the public
        \end{itemize} & \vspace{-1\baselineskip}\begin{itemize}[leftmargin=*]
            \item Infected users' identities $\rightarrow$
            
            \{tech-savvy users\}
        \end{itemize} \\
    \bottomrule
    \end{tabular}
    \label{tab:six_app_design_options}
\end{table}

% We aimed to make sure that although participants will not see it, each app option can be backed up with a feasible implementation.

\label{sec:why_centralized_identity}
Note that although there are apps that use a centralized architecture and do not require every user to register with their real identities, we did not include them in our survey and always consider centralized apps allow central authorities to know all users' identities.
One reason is that the central server already collects sufficient data to build a social graph, which makes it possible to re-identify people \cite{potterat2002risk}.
Another reason is that malicious users may open multiple accounts to help them narrow down the identities of infected users that they have been exposed to, which negates the benefits of centralized designs, namely not disclosing infected users' identities to other users, as discussed in prior work \cite{GitHubDP97:online}.

\subsection{Survey items}

We describe the five sections of the survey below.
Our complete survey is available at \url{https://git.io/Jfz61}.

\paragraph{Section 1: Study introduction and consent form} This part gives participants a brief overview of our study, and requests them to read and sign the consent form if they want to proceed.

\paragraph{Section 2: Background about COVID-19 and contact-tracing apps} To help our participants make informed decisions based on a full comprehension of the risks and benefits with contact-tracing apps, we begin our survey with a brief introduction about COVID-19 and contact-tracing apps.
We used three simulation videos to illustrate how fast the disease spread in three conditions: without any intervention, with a perfectly accurate contact-tracing app that achieved 100\% adoption rate, and with a perfectly accurate contact-tracing app that only achieved 20\% adoption rate\footnote{These videos can be watched using the following links: \url{https://youtu.be/tP8h9FpuFFY} (no intervention condition),  \url{https://youtu.be/8yE1Sf5HhPw} (perfect contact tracing app condition), \url{https://youtu.be/4Td8pwOBppY} (partial adoption of contact tracing app condition)}.

The format of the simulation videos are inspired by an article published on the Washington Post \cite{Whyoutbr53:online}.
We modified an open-sourced implementation of their simulation and also open-sourced our implementation \footnote{\url{https://github.com/covid19-hcct/covid-19-spread-simulator/releases/tag/survey\_2020May}}.
% no intervention: https://youtu.be/tP8h9FpuFFY
% full contact tracing: https://youtu.be/8yE1Sf5HhPw
% partial contact tracing: https://youtu.be/4Td8pwOBppY

\begin{figure}[h]
    \centering
    \includegraphics[width=0.6\linewidth]{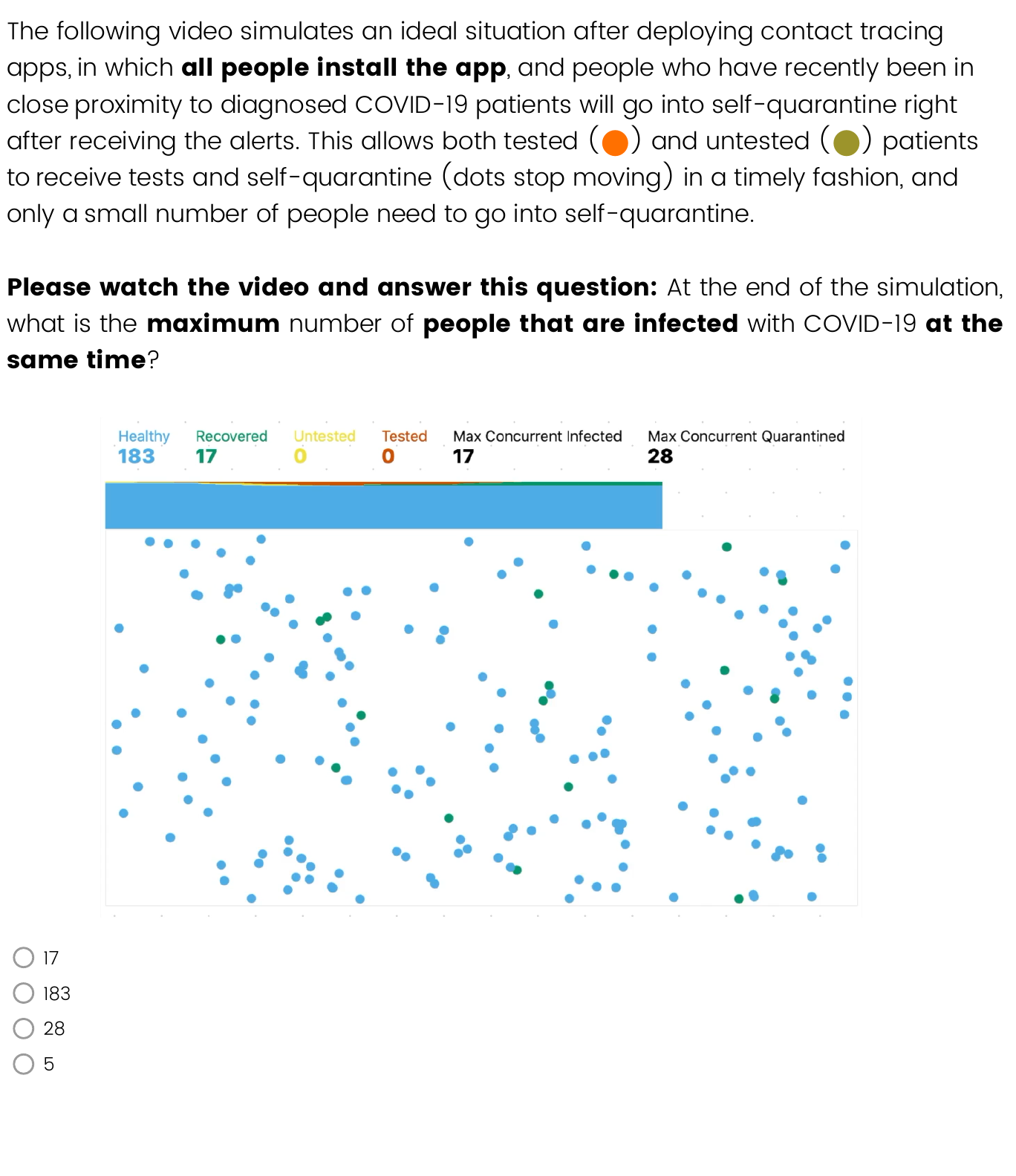}
    \caption{Survey Section 2: An example of the simulation video and the corresponding attention check question in the survey. The correct answer will be the value of ``Max Concurrent Infected'' at the end of the video.}
    \label{fig:background_attentioncheck_screenshot}
\end{figure}

Our simulation presents the spread of COVID-19 among people (represented by dots) in a fixed area for a certain period of time (Figure \ref{fig:background_attentioncheck_screenshot}).
Each dot has five possible \emph{status}: \emph{Healthy}, \emph{Recovered}, \emph{Exposed}, \emph{Infected (untested)}, \emph{Infected (tested/diagnosed)}.
The simulation starts with a few people infected with the disease but not diagnosed (\emph{Infected (untested)}).
They will be diagnosed over time and be marked as \emph{Infected (tested)}, and they will cease to move to simulate a quarantined state.
\emph{Infected (untested)} will expose every healthy person (\emph{Healthy}) they contacted to become \emph{Exposed}.
\emph{Exposed} people have a chance to become infected (\emph{Infected (untested)}).
And finally, \emph{Infected (untested)} and \emph{Infected (tested/diagnosed)} people will recover (\emph{Recovered}), and will not be infectious.

For contact tracing conditions, depending on the adoption rate (100\% in the second video and 20\% in the third video), a proportion of in-person contact with be recorded, so when a person is diagnosed and turn \emph{Infected (tested)}, their close contacts that have been recorded will be quarantined immediately, including those who had not been infected (\emph{Exposed}) or infected but not tested(\emph{Infected (untested)}).

As the simulation goes on, the current number of people in each status will be presented at the top and updated in real time. The video also shows what is the maximum number of people that are infected/quarantined at the same time. As shown in Figure \ref{fig:background_attentioncheck_screenshot}, we asked a simple question regarding the numbers shown in the video as attention check questions, such as ``What is the maximum number of concurrently infected people?''. These questions have only one correct answer, and are a common technique in survey studies to check whether the respondents are paying attention to the survey. We later removed all responses that did not correctly answer all three attention check questions when analyzing the data.

\begin{figure}[h!]
    \centering
    \includegraphics[width=0.9\linewidth]{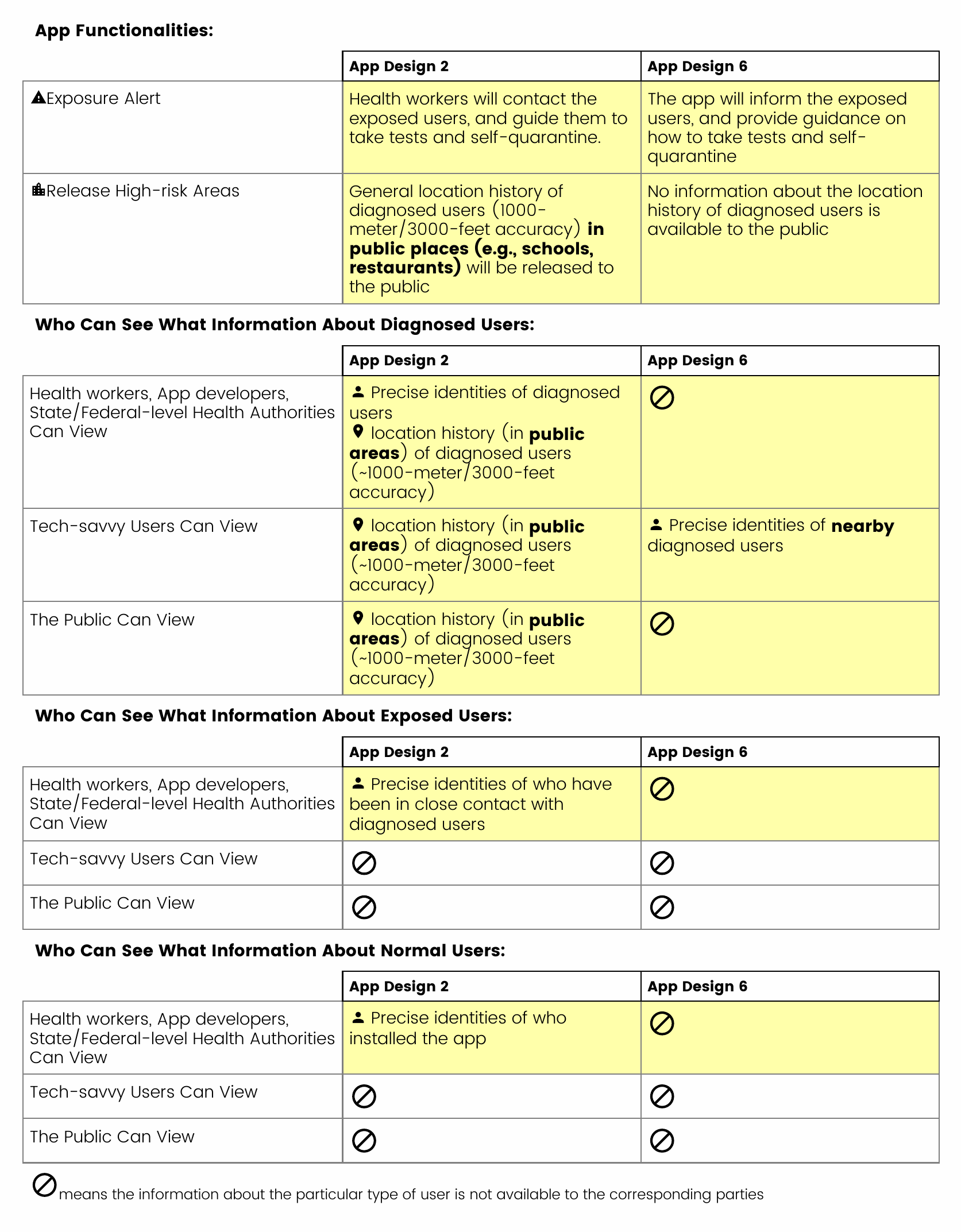}
    \caption{Survey Section 3: An example of the visuals we used in our survey to help participants compare two app designs. Differences between the two design are highlighted in light yellow.}
    \label{fig:pairwise_comparison_viz}
\end{figure}

\paragraph{Section 3: Pairwise comparison of app design options: relative installation preferences} 
This part asks participants to compare app designs in pairs and explain their choices in free-form responses to help us understand which design option was preferred by more users and why.
Figure \ref{fig:pairwise_comparison_viz} demonstrates how we present the pairs in our survey.
All participants were presented with all possible combinations of the six app designs (15 pairs).
The overall order of the pairs were randomized for each participant.
To minimize potential bias, we did not include any text implying the usefulness of any specific functionality or the privacy sensitivity of any specific data practice. 

Since it is easier for people to compare options in pair than rate them individually, we included these pairwise comparison questions to allow our participants to gain a thorough understanding in the six designs.
However, pairwise comparison can only elicit relative judgment, which does not precisely reflect people's intention to install the app, especially when multiple options were perceived as equally good or bad. Therefore, we later asked our participants to rate the six designs individually in the Section 5 of the survey.

\begin{figure}[h!]
    \centering
    \includegraphics[width=0.7\linewidth]{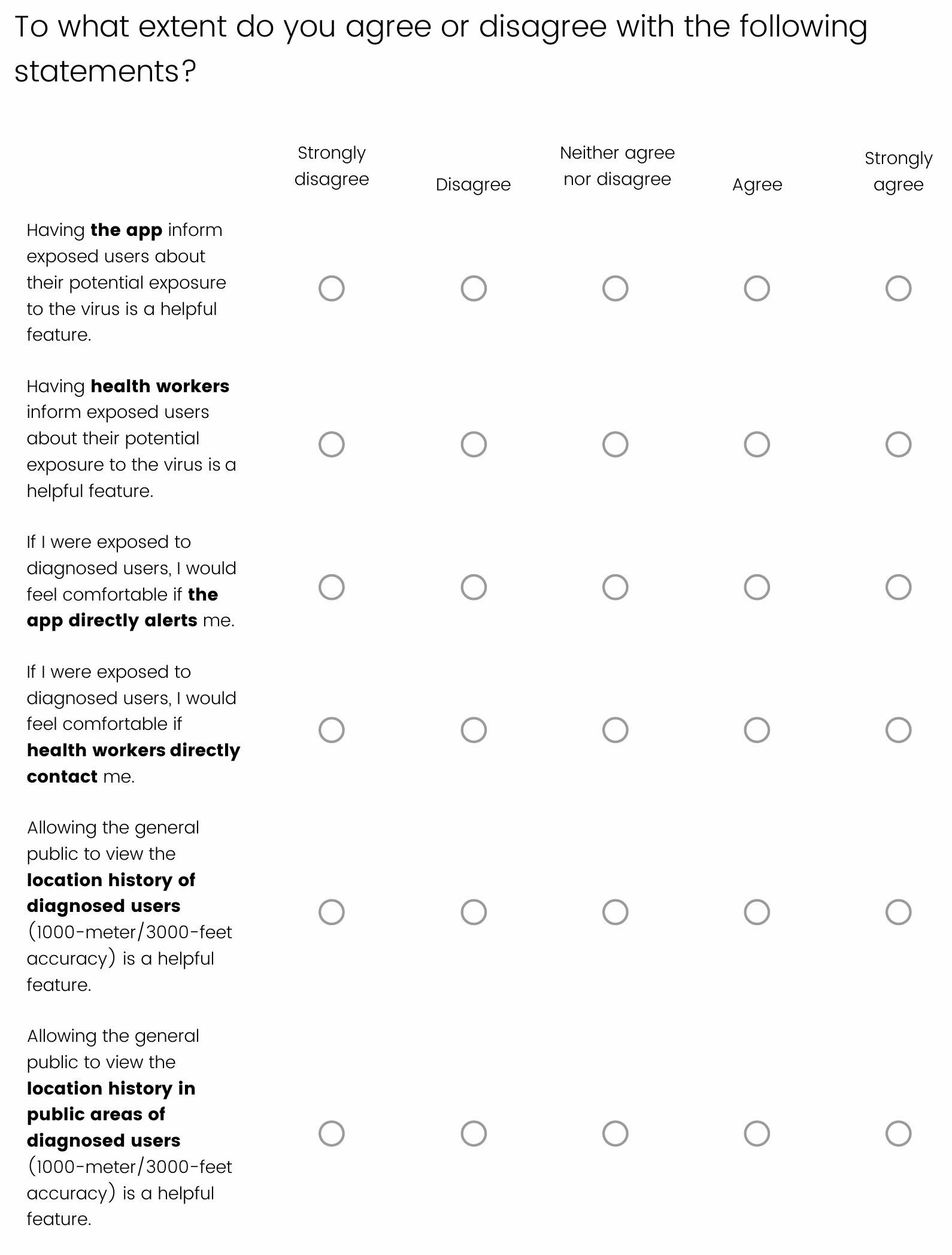}
    \caption{Survey Section 4: An example of how we presented the individual factor rating questions in our survey. Similar statements are grouped together with the differences highlighted in bold. Participants were asked to rate to what extent they agree or disagree with these statements (5-point Likert scale).}
    \label{fig:individual_usefulness_screenshot}
\end{figure}

\paragraph{Section 4: Individual factor rating} Because multiple factors (e.g., data sharing practices, app functionality) could affect users' feelings about one app design, we then asked participants how they feel about these factors individually. 
Specifically, we presented a set of statements regarding privacy and app functionality, and asked participants to rate to what extent they agree or disagree with these statements (5-point Likert scale). Figure \ref{fig:individual_usefulness_screenshot} shows a screenshot of part of this section of the survey to demonstrate how we presented the statements.

There are 30 statements regarding privacy divided into three groups.
In this part, we prompted participants to think about data sharing practices from the perspectives of users who are in a certain status, including infected (3 types of data), exposed to the virus (1 type of data), and normal (i.e., not infected or exposed and just installed the app, 1 type of data.)
For each type of data in a certain status, we listed six statements, each corresponding to sharing the data with a specific stakeholder mentioned in Section \ref{sec:six-stakeholders}.
Because among all six app designs, only the location information of diagnosed users will be disclosed to any of the stakeholders, we had 12 statements regarding location data sharing from the perspectives of infected users, including both options of sharing all locations and public locations.
The other 18 statements are about identity information sharing, from the perspectives of all three users status.
Then format of the statements is ``My [data type] must be kept private from [stakeholder type].''

There are 6 statements regarding the app functionality. The first four statements are regarding the usefulness and comfort level of different approaches to provide exposure notices (health workers vs. app informing), and the other two statements are regarding the perceived usefulness of hotspot information derived from all locations vs. public locations. Figure \ref{fig:individual_usefulness_screenshot} shows all the six statements.

\paragraph{Section 5: Willingness-to-install ratings: Absolute app installation preferences}
In this section, we present all six app designs again on one page, laying them side by side, and ask participants to rate how strongly they agree or disagree (5-point Likert scale) with the statement ``If [app id] were available, I would install it on my phone'' for each app design option.
This section is designed to help us understand people's absolute willingness to install apps using each of the six designs. We asked these questions after the pairwise comparison questions, so participants had familiarized themselves with the six designs during the pairwise comparison process, which makes the absolute rating easier.

\paragraph{Section 6: Demographic information} In this section, we collect the demographic information such as age, gender, education, household income, which state they currently reside in, and ethnicity. We also asked questions related to the current status, such as their attitudes towards COVID-19, how their life can be affected by stay-at-home orders, how much they trust different levels of governments (e.g., city/county, state, federal), and their privacy attitudes.

\subsection{Study procedure}

We recruited 244 people from Amazon Mechanical Turk from April 27-May 7, 2020. Amazon Mechanical Turk (MTurk) is an online crowdsourcing platform designed to help recruit people to achieve various types of tasks. It is a common platform for survey studies due to the capability of gathering a diverse sample in a short time frame~\cite{paolacci2010running}. We used the same sampling criteria as in previous studies to increase quality \cite{kelley2010conducting}, by restricting the participants to those with an approval rate of at least 95\%. We also restricted the participants to be in the U.S. Each participant was paid \$3 for completing the survey. The survey takes about 20 minutes to complete.

For the collected responses, we first removed 16 responses that did not pass all three attention check questions. We then removed 20 responses whose geolocated IP address (i.e. the state) did not match what the MTurk 
worker self-reported state in the U.S. Note that we manually corrected all mismatches caused due to the IP address being located at the border of two neighboring states. We also checked the MTurk ID to avoid counting multiple responses from the same person. The study has been approved by the Institutional Review Board of Carnegie Mellon University.

\section{Results}

% What is our sample like: demographic descriptive stats
Our sample has 208 participants from 41 states in the United States. Table \ref{tab:sample_overview} summarizes the demographic characteristics of our sample.

% Result-below bachelor:Coefficients:
%                  Estimate Std. Error z value Pr(>|z|)
% (Intercept)       -0.9725     0.1844  -5.275 1.33e-07 ***
% centralization1    0.6103     0.1762   3.463 0.000533 ***
% public_location1   0.5561     0.2168   2.565 0.010326 *
% all_location1      0.3535     0.2178   1.623 0.104499

% Result-bachelor or above:
% Coefficients:
%                  Estimate Std. Error z value Pr(>|z|)
% (Intercept)       -1.0584     0.1666  -6.352 2.12e-10 ***
% centralization1    0.7142     0.1589   4.495 6.95e-06 ***
% public_location1   0.5766     0.1944   2.966  0.00302 **
% all_location1      0.2121     0.1966   1.079  0.28067

\begin{table}[h]
    \centering
    \caption{Demographic characteristics of our survey sample collected on MTurk. total $N=208$}
    \begin{tabular}{p{0.4\linewidth}R{0.1\linewidth}R{0.1\linewidth}}
    \toprule
    Demographic Characteristics & N & Percentage \\
    \midrule
    \textbf{Gender}\\
    \hspace{3mm} Female & 118 & 56.7\%\\
    \hspace{3mm} Male & 88 & 42.3\%\\
    \hspace{3mm} Other & 1 & 0.5\% \\
    \hspace{3mm} Prefer not to say & 1 & 0.5\%\\
    \textbf{Age}\\
    \hspace{3mm}18--24 & 1 & 0.5\%\\
    \hspace{3mm}25--34 & 52 & 25.0\%\\
    \hspace{3mm}35--44 & 73 & 35.1\%\\
    \hspace{3mm}45--54 & 39 & 18.8\%\\
    \hspace{3mm}55--64 & 30 & 14.4\%\\
    \hspace{3mm}65+ & 13 & 6.3\%\\
    \hspace{3mm} \small{(Mean age: 43.4, Min age: 24, Max age: 73)} \\
    \textbf{Education}\\
    \hspace{3mm} Below bachelor's degree & 92 & 44.2\% \\
    \hspace{3mm} Bachelor or above bachelor's degree & 116 & 55.8\% \\
    \textbf{Last election voting results (2016)} \\
    \hspace{3mm} From states voted Democrat & 96 & 46.2\% \\
    \hspace{3mm} From states voted Republican & 112 & 53.8\% \\
    \bottomrule
    \end{tabular}
    \label{tab:sample_overview}
\end{table}

\subsection{Overall preferences: Apps that follow centralized designs and release public-area hotspots are significantly more likely to be installed at country level}
\label{sec:overall_preference_results}

\begin{figure}[h]
    \centering
    \includegraphics[width=0.8\linewidth]{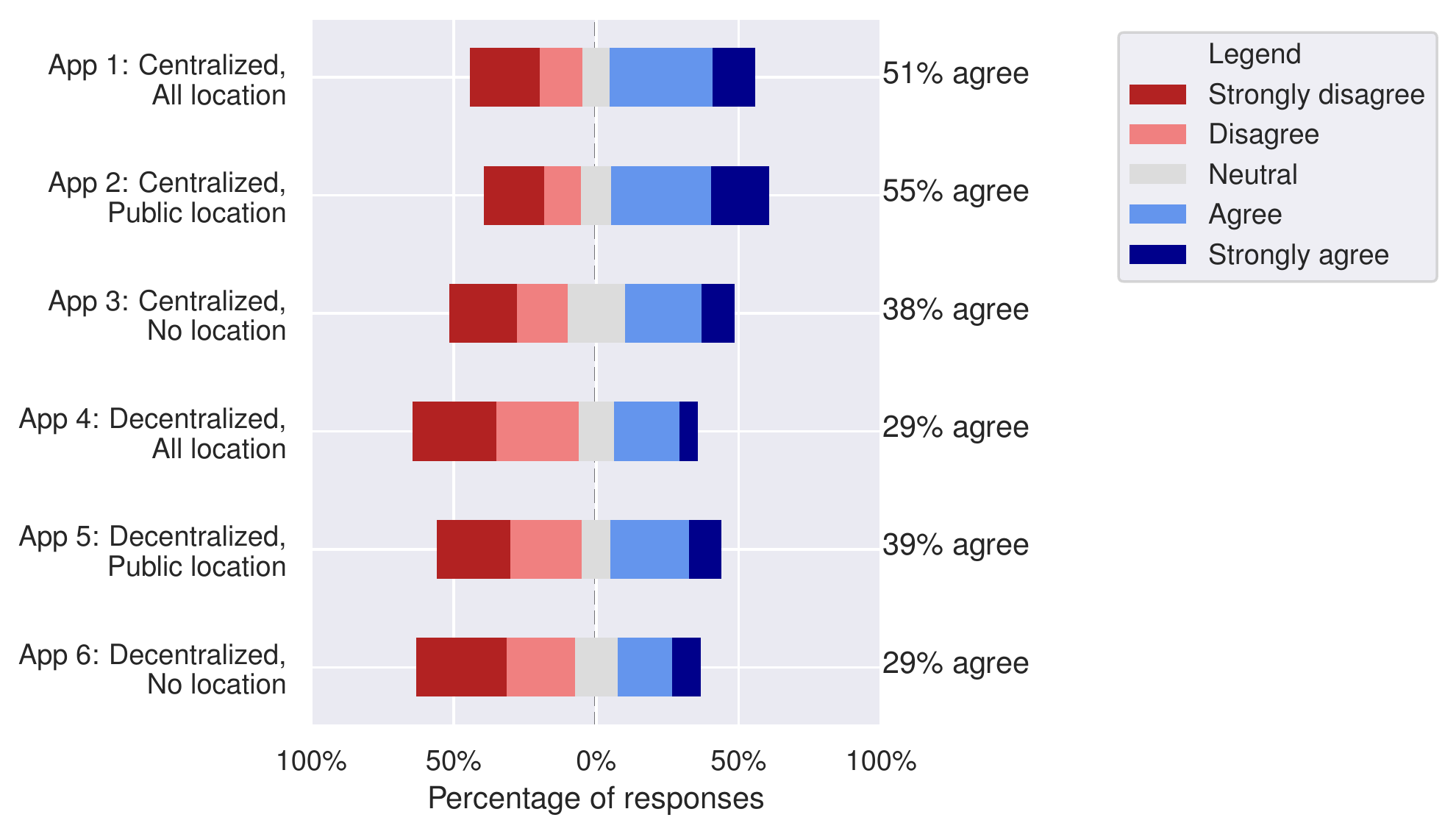}
    \caption{Overall willingness to install app (5-point agreement Likert scale).
    The statement is "If [app id] is available, I would install it on my phone."
    The app design option that has the highest percentage of participants agree to install is App 2, following a centralized design and sharing public-area location history from infected users with the public.}
    \label{fig:app_install_rate_overall}
\end{figure}

Figure \ref{fig:app_install_rate_overall} shows the results of our participants' self-reported willingness to install apps based on the six different designs (Survey Section 5). Among the six options, App 2 (Centralized + public locations)
was the most popular, with a total of 55\% participants strongly agreeing or agreeing to install. The least popular app was App 6, with only 29\% strongly agreeing or agreeing to install. Note that the two most popular designs (App 1 and App 2) both had more than 50\% of participants self-reporting their willingness to install, a result that is close to a survey conducted by
UMD and the Washington Post \cite{Washingt23:online} asking people about installing a general contact-tracing app (50\%).

% \jason{For Fig 1, clarify what the percentages are in the left and right side. Also, suggest using a different font so that people don't immediately go from (say) App 6 to 56\%, need to make the columns of numbers more distinct}

We conducted a logistic regression analysis to further investigate the correlation between the design choices and participants' preferences. The independent variable is the app choices, and the dependent variable is whether to install the app (Strongly agree or agree = 1; Other = 0.) The results are presented in Table \ref{tab:logistic_regression_install}. In the first model (Row 1), we only differentiate between collecting or not collecting location, but not between different levels of location granularity. The results show that, in general, users are significantly more likely to install the app if it follows a centralized architecture (as compared to decentralized architecture), and provide hotspot information (as compared to not collecting location).

In the second model (Row 2), we added another dummy variable to separate all location history design from public-only location history design.
The results show that users are significantly more likely to install the app if it provides information about hotspots in public areas only (``public location only'') as compared to not collecting location.
The dummy variable for ``All location'' does not show significant effect.
This suggests that collecting location data in public areas only is potentially a good way to balance the privacy and utility requirements.

We then explored whether these preferences hold in different regions in the U.S., which will matter if state-specific solutions are adopted (e.g., the app ``Care19'' in North Dakota~\cite{Care19ND41:online}). A recent study by the Washington Post suggested that one's partisan leaning could affect people's willingness to install contact-tracing apps in general, and identified that people who self-identified as Republicans were less amenable to install contact-tracing apps~\cite{Washingt23:online}.
In a similar spirit, we separate our responses into two groups based on the 2016 United States presidential election results of the states they currently reside in.
Although centralized designs and public location only achieved significant positive effect on willingness to install the app, the corresponding effect sizes (odds ratio, abbreviated as OR in Table \ref{tab:logistic_regression_install}) of the Blue states are consistently larger than the Red states, which suggests that the most suitable app design in different states may vary according to partisan leanings.
Odds ratio (OR) quantifies the strength of the association between the dependent variable (willingness to install) and the independent variable (design choices).
% Odds means the ratio of the estimated probability of installing the app to not installing the app, and \emph{odds ratio} means the ratio of the odds when the dependent variable equals to 1 (e.g., choosing the Centralized option in the Centralized vs. Decentralized comparison) to the odds when the dependent variable equals to 0 (e.g. choosing Decentralized).
If OR is larger than 1, it means there is a \emph{positive effect} on willingness to install of this design choice, and the larger the OR the greater the effect; If OR is smaller than 1, it means there is a \emph{negative effect} on willingness to install of this design choice, and the smaller the OR the greater the effect.
Our results suggest that the centralized design option has a smaller advantage in states that lean Republican than states that lean Democratic.

% \jason{For Table 4 caption, keep in mind that many readers may not be familiar with the columns, suggest adding either briefly in caption but more in the text how to interpret things. }

% \jason{Skimming ahead, it's not clear to me if other analyses were done, e.g. education level, attitudes toward COVID, etc. Basically, the demographics part of the survey seems like there are many opportunities for analysis.}

\begin{table}[h]
    \centering
    \caption{Logistic regression between app install preferences and app design choices. The dependent variable is whether to install the app (Strongly agree or agree = 1, Other = 0), and the independent variables are the app choices (e.g., centralized vs. decentralized, collect all location vs. collect location in public areas vs. do not collect location.) The first two models suggest that centralized designs and collecting location in public areas are better choices at country level that can result in higher installation rates. We then separate participants into two groups based on who their states voted for in the 2016 United States presidential election. Although centralized designs and public location only achieved significant positive effect on willingness to install the app, the corresponding effect sizes (OR) of the Blue state group are consistently larger than the Red states, which suggests that the most suitable app design at state level may vary according to the partisan leanings.}

    \begin{tabular}{p{0.1\linewidth}p{0.33\linewidth}R{0.07\linewidth}p{0.06\linewidth}R{0.08\linewidth}p{0.02\linewidth}p{0.1\linewidth}p{0.05\linewidth}}
        \toprule
        Model & Predictor & $\beta$ & $SE$ $\beta$ & Wald's $\chi$ & $df$ & $p$ & $OR$\\
        \midrule
        1-All & (Intercept) &  -1.0184   &  0.1235 & 68.1 & 1 & <.001 *** & 0.3612 \\
        &Centralized (1) vs. Decentralized (0) \\
        &\hspace{3mm} Centralized & 0.6649 & 0.1177 & 31.9 & 1 & <.001 *** & 1.9444 \\
        &Location use (no location = 0, public/all location = 1)\\
        &\hspace{3mm} Collect location and release hotspots & 0.4226 & 0.1266 & 11.1 & 1 & <.001 *** & 1.5260\\
        \midrule
        2-All & (Intercept) &  -1.0197   &  0.1236 & 68.1 & 1 & <.001 *** & 0.3607 \\
        & Centralized (1) vs. Decentralized (0) \\
        &\hspace{3mm} Centralized & 0.6673 & 0.1179 & 32.0 & 1 & <.001 *** & 1.9490 \\
        &Location use (no location = 0)\\
        &\hspace{3mm} Public location only & 0.5673 & 0.1447 & 15.4 & 1 & <.001 *** & 1.7635\\
        &\hspace{3mm} All location & 0.2755 & 0.1459 & 3.6 & 1 & .0589 & 1.3172 \\
        \midrule
        \multirow{2}{*}{2-Red states} & (Intercept) &  -0.8611   &  0.1645 & 27.4 & 1 & <.001 *** & 0.4227 \\
        & Centralized (1) vs. Decentralized (0) \\
        &\hspace{3mm} Centralized & 0.5998 & 0.1594 & 14.2 & 1 & <.001 *** & 1.8217 \\
        &Location use (no location = 0)\\
        &\hspace{3mm} Public location only & 0.4332 & 0.1949 & 4.9 & 1 & .0262 * & 1.5422\\
        &\hspace{3mm} All location & 0.1353 & 0.1967 & 0.47 & 1 & .4916 & 1.1449 \\
        \midrule
        \multirow{2}{*}{2-Blue states} & (Intercept) &  -1.2155   &  0.1880 & 41.8 & 1 & <.001 *** & 0.2966 \\
        & Centralized (1) vs. Decentralized (0) \\
        &\hspace{3mm} Centralized & 0.7507 & 0.1757 & 18.3 & 1 & <.001 *** & 2.1184 \\
        &Location use (no location = 0)\\
        &\hspace{3mm} Public location only & 0.7322 & 0.2167 & 11.4 & 1 & <.001 *** & 2.0796\\
        &\hspace{3mm} All location & 0.4473 & 0.2181 & 4.2 & 1 & .0403 * & 1.5641 \\

        \bottomrule
    \end{tabular}
    \label{tab:logistic_regression_install}
\end{table}

\subsection{Understanding what factors affect people's preferences using ratings about each individual factor and the pairwise comparison results}

Although previous results have suggested that people prefer centralized solutions and prefer apps that provide them with hotspot information despite the cost of location privacy, we do not yet fully understand what factors contribute to this difference. This section aims to answer this question by analyzing the ratings of individual factors and the free-form explanations for pairwise app design comparisons.

\subsubsection{Analyzing individual privacy-related factor ratings: Identities are more sensitive than location, and tech-savvy users accessing sensitive data is more concerning than health authorities}
\label{sec:privacy_related_factors}

\begin{figure}[h]
    \centering
    \includegraphics[width=0.8\linewidth]{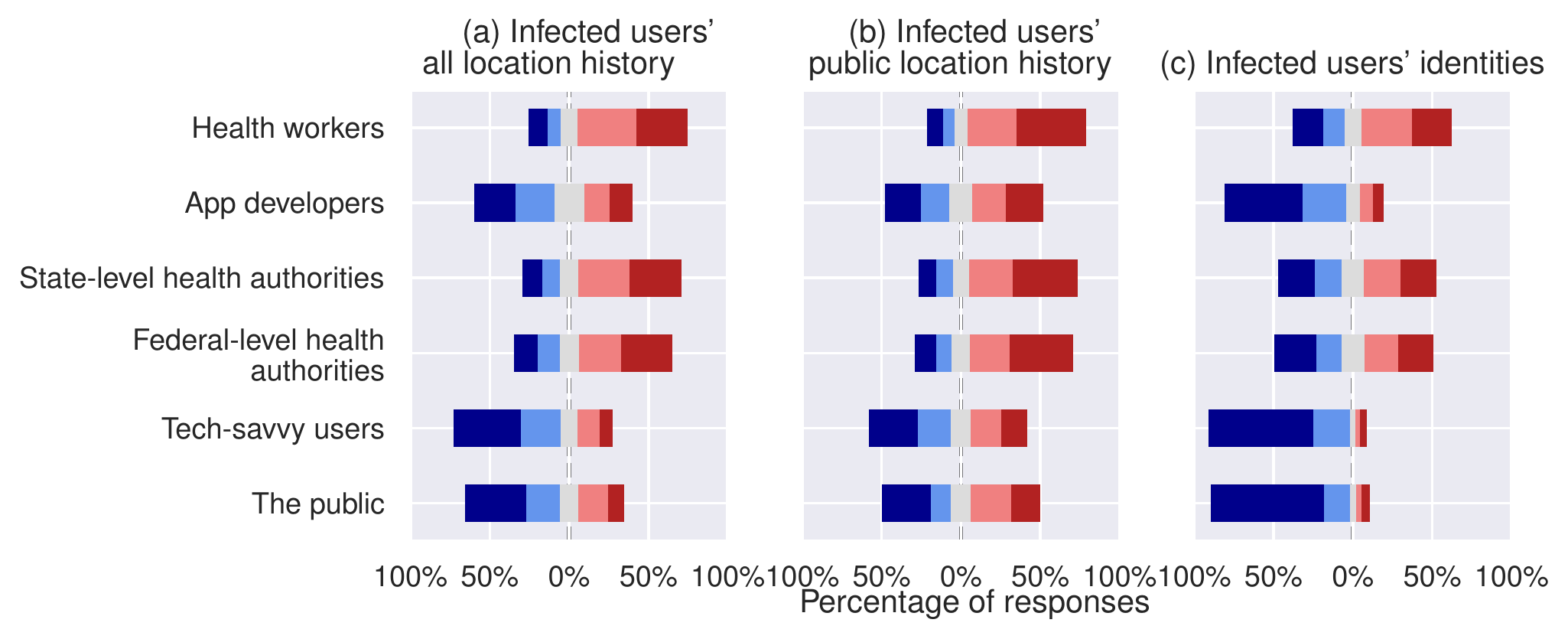}
    \includegraphics[width=0.8\linewidth]{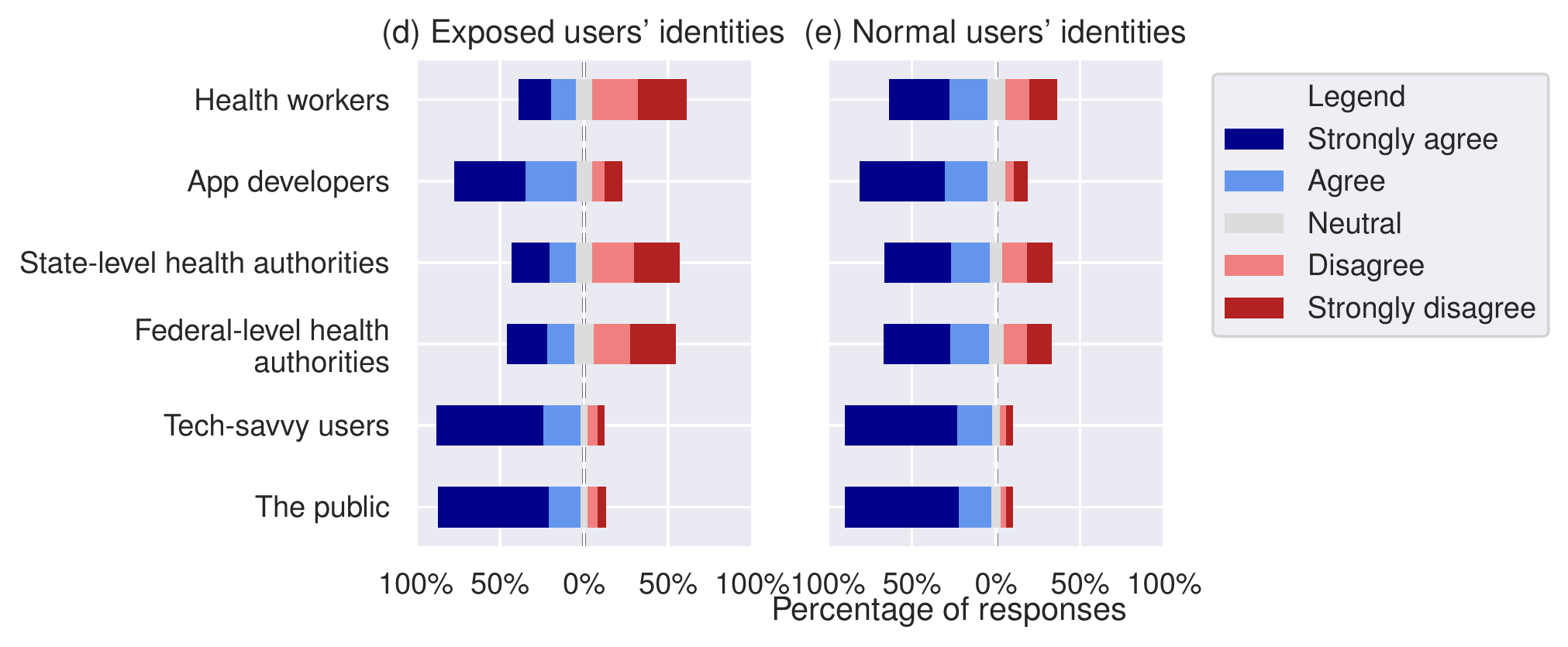}
    \caption{This plot shows to what extent our participants agreed or disagreed with the statements ``If I were [infected/exposed/normal], my [all location history/public location history/identities] must be kept private from [one of the six stakeholders]''. For example, ``If I were \emph{infected} with COVID-19, my \emph{real identity} must be kept private from \emph{health workers}.'' Our result suggests that 1) People find sharing sensitive data with health workers and health authorities (both state and federal levels) more acceptable than sharing the same data with app developers and other users. 2) People in general are more concerned about sharing their real identities than sharing coarse-grained location history (e.g., \char`\~1000m/3000ft accuracy.) 3) People who have been diagnosed with COVID-19 or have been exposed to the virus may find sharing data more acceptable than normal users.}
    \label{fig:individual_likert_privacy_plot}
\end{figure}

Figure \ref{fig:individual_likert_privacy_plot} presents the 5-point Likert-scale responses to statements ``If I were [infected/exposed/normal], my [all location history/public location history/identities] must be kept private from [one of the six stakeholders]'' to understand how concerned people are about different data sharing practices. The longer the bars show up in the right half of the plot, the more comfortable people feel about sharing the data to the specific target.
All statistical test results below are under two-sided Mann-Whitney U tests.

We first analyzed the sensitivity of different sharing targets. Figure \ref{fig:individual_likert_privacy_plot} shows a trend that people are generally more comfortable with having their location data and identity information accessible by health workers, state-level health authorities, and federal-level health authorities (with more than half agreed to share), and less comfortable with having the same data accessible by app developers, tech-savvy users and the public (with more than half disagree to share). We also conducted statistical analysis to verify the above observations. Specifically, we tested the difference between ratings related to sharing data with tech-savvy users and ratings related to sharing data with state/federal-level health authorities, and observed a significant difference ($U=1556559.5$, $p<.001$.)
This shows that our participants were generally more concerned about tech-savvy users inferring their identities (happens in a decentralized solution~\cite{GitHubDP97:online}) than health authorities (happens in a centralized solution), which explains why more people preferred to install apps that followed centralized design.
We also tested the difference between ratings related to sharing data with state-level health authorities and ratings related to sharing data with federal-level authorities, and did not get a statistically significant result ($U=527568.5$, $p=0.32$.)

We also observed that people seem to feel more concerned about sharing their identities than sharing their location data, and feel more comfortable sharing public location history than all location history if they were infected. The difference between all ratings related to sharing location and sharing identities ($U=1122758.5$, $p=<.001$) and the difference between sharing public location history and all location history ($U=871770.0$, $p<.001$) are both statistically significant.

Lastly, we compared people's feelings about the privacy risks under different situations. Specifically, we compared to what extent they wanted to keep their identities private under the three conditions: infected (Figure \ref{fig:individual_likert_privacy_plot}a-c), exposed (Figure \ref{fig:individual_likert_privacy_plot}d), and normal (not infected or exposed and just installed the app, Figure \ref{fig:individual_likert_privacy_plot}e.) We observed that normal users tend to have more concerns about disclosing their identities (i.e., allowing others to know that they installed the app). The difference is statistically significant ($U=661852.5$, $p<.001$.)
Similarly, normal users also seemed to be more concerned about sharing their identities than infected users, and the difference is statistically significant ($U=692459.0$, $p<.001$.)

\subsubsection{Analyzing individual usefulness-related factor ratings: All features were perceived as useful, and releasing public location history of infected users was perceived as more useful than releasing all location history.}
\label{sec:utility_related_factors}

\begin{figure}[h]
    \centering
    \includegraphics[width=0.8\linewidth]{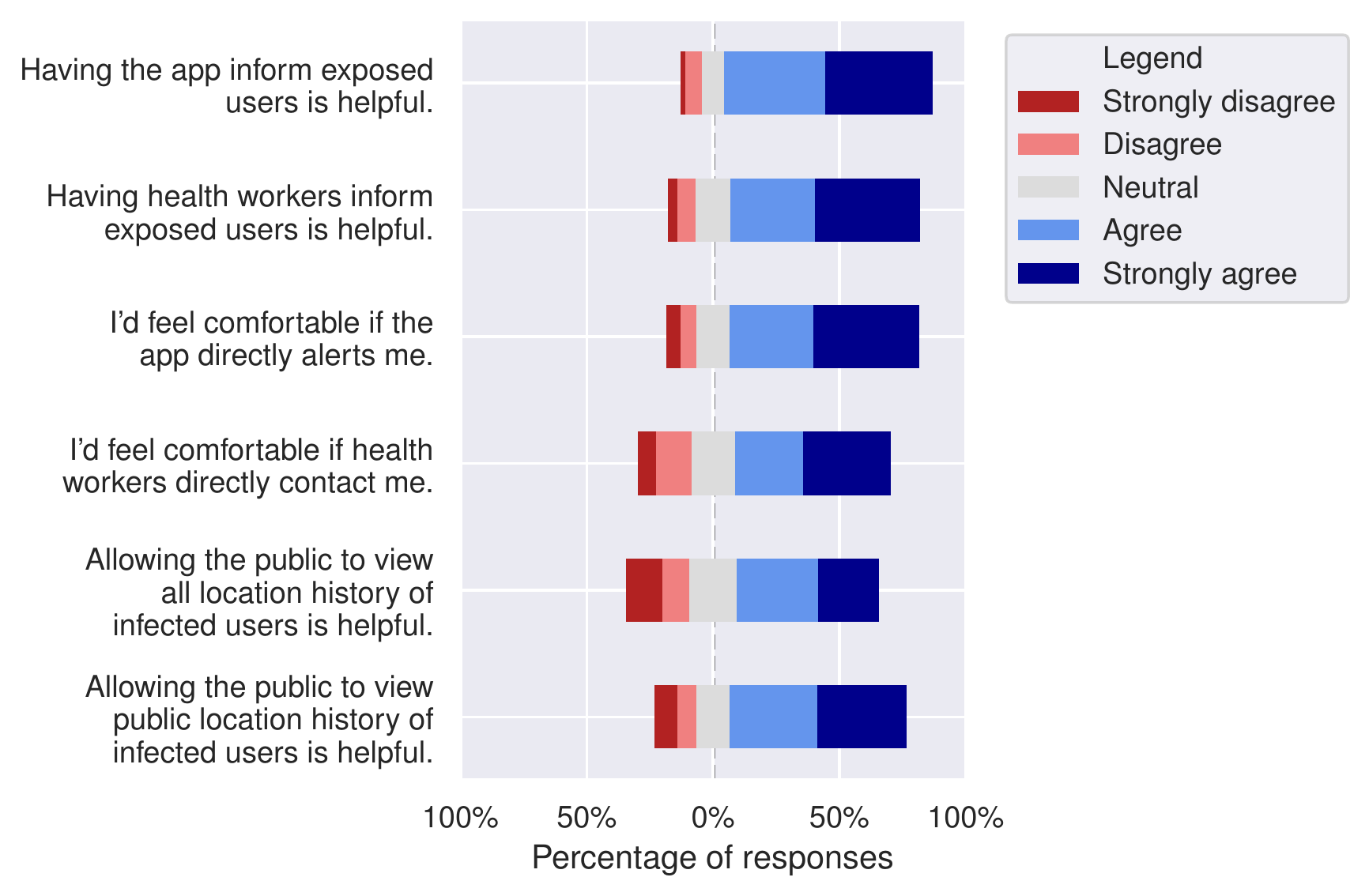}
    \caption{This plot shows to what extent our participants agreed or disagreed with the statements on the left. Our results showed that both the basic contact tracing and hotspot features were perceived as useful. People were significantly more comfortable with the app informing them of potential exposure, and perceived information about hotspots in public areas as more useful.}
    \label{fig:individual_likert_usefulness_plot}
\end{figure}

Figure \ref{fig:individual_likert_usefulness_plot} shows participants' 5-point agreement Likert-scale responses to statements related to app features (collected in Section 4 of the survey). All statistical test results below are under two-sided Mann-Whitney U tests.

The first four statements are related to the exposure notification methods that can be affected by choosing centralized or decentralized design. The results regarding the first two statements showed that 83\% and 75\% of our participants agreed that the two features are helpful, respectively. Although there is no significant difference between the first two statements ($U=22728.0$, $p=0.34$), the difference between the third and the fourth statement is significant under the same test ($U=24563.5$, $p=0.012$.) This suggests that people may feel more comfortable if the app directly alerts them than real human beings contacting them.

The last two statements are related to providing the public with more information about the recent whereabouts of infected users. Interestingly, although only sharing location history in public areas provide less information than in all places, it was perceived as more useful by our participants (70\% agree or strongly agree than 56\%). This difference is statistically significant ($U=17886.5$, $p=0.0015$.)

\subsubsection{Analyzing free-form responses: The privacy-utility trade-off in the use of location data, and the choices between sharing identity data with the authorities or tech-savvy users are two salient themes}
\label{sec:qualitative_codes}

We then looked into the free-form responses explaining participants' preferences in the pairwise comparison tasks to better understand what factors they mentioned when explaining why they prefer one app design over another.

\paragraph{Coding process}
Our analysis was conducted on 750 responses of 50 participants randomly selected from the 208 participants. Two researchers (the first and second author) first conducted open coding~\cite{khandkar2009open} on the data collectively. 35 codes that captured reasons emerged in the text were developed during this process. These two researchers then conducted axial coding to group these codes into high-level categories, resulting in a framework that contains four main categories and 14 codes. The 750 responses were then coded again with the final coding framework.

\paragraph{Coding results}

\begin{figure}[h]
    \centering
    \includegraphics[width=0.6\linewidth]{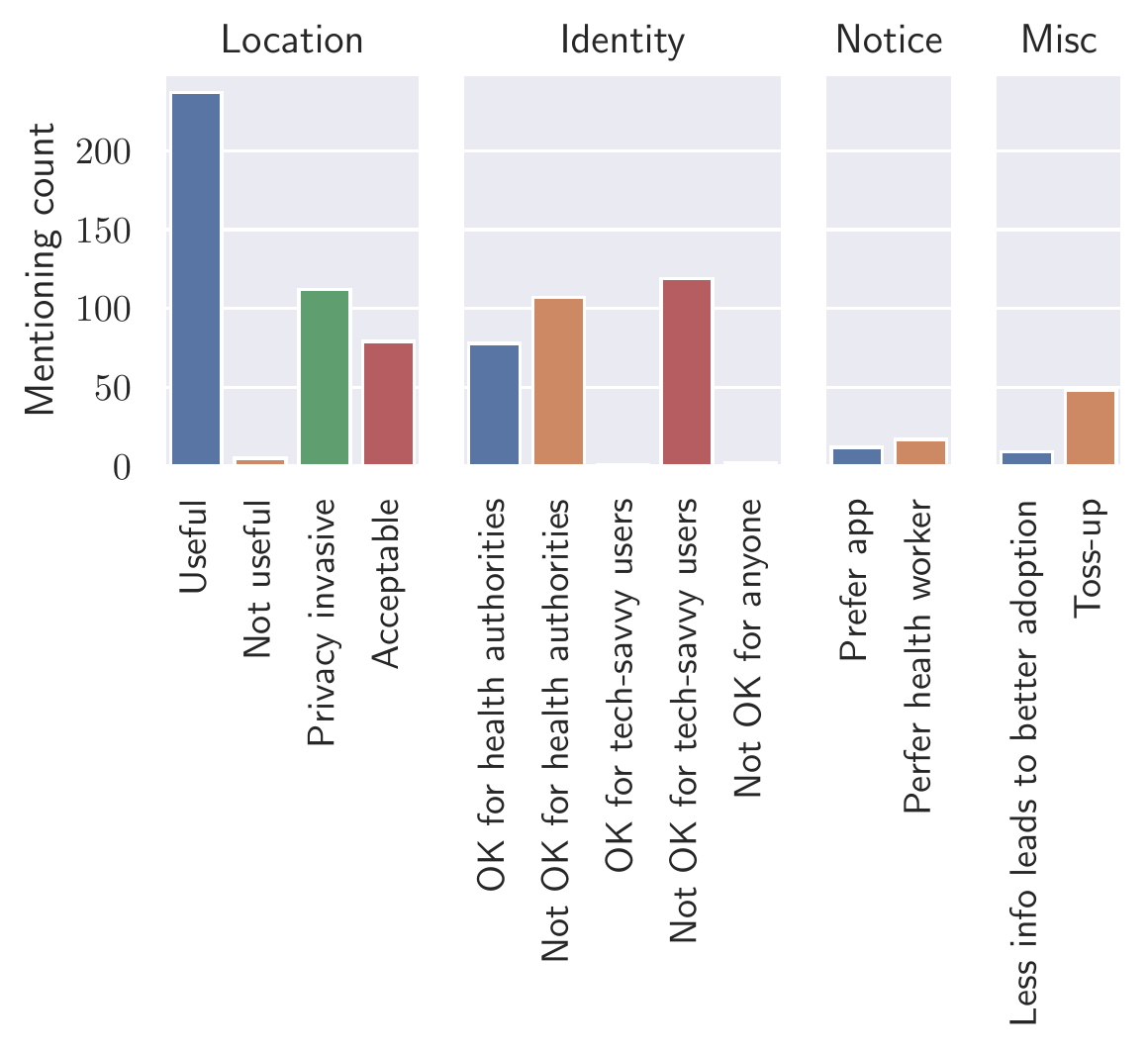}
    \caption{This figure demonstrates the mentioning counts of codes that capture factors people ascribed to when choosing which app design they preferred. Regarding comparisons with different location-based features, this result shows that both utility and privacy are important factors that affect people's choices. However, when it comes to the comparisons between centralized and decentralized designs, the privacy aspects (i.e., sharing identity information with whom.) were a lot more frequently discussed than the utility aspect (i.e., how the exposure notice is delivered.). In addition, almost an equivalent number of responses mentioned concerns about sharing identity information with authorities and tech-savvy users, which suggests that neither of the centralized and decentralized design could address public privacy concerns perfectly.}
    \label{fig:all_code}
\end{figure}

Table \ref{tab:all_code_examples} summarizes the final framework, and Figure \ref{fig:all_code} demonstrates how many responses mentioned these codes in the labeled sample. The first three high-level categories happen to correspond with the potential privacy-utility trade-offs in our six designs.

\begin{table}[h!]
    \centering
    \caption{This table summarizes factors that explain people's preferences about app designs that emerged in the free-form responses. The first three categories happen to match the privacy-utility trade-offs in the six app designs. This qualitative analysis allows us to understand why people preferred certain designs over others. For example, the choices between centralized and decentralized solutions seemed to be more affected by privacy concerns (i.e., whether sharing identity information with authorities or tech-savvy users is more acceptable) than utility (i.e., which method to inform exposed users is more helpful.)}
    \begingroup
    \setlength{\tabcolsep}{7pt} % Default value: 6pt
    \renewcommand{\arraystretch}{1.5} % Default value: 1
    \begin{tabular}{p{0.15\linewidth} p{0.3\linewidth} p{0.4\linewidth}}
    \toprule
    Category & Code & Example quotes \\
    \midrule
    Location & Useful & \textit{Not sure I'm comfortable with App 1, but app 3 seems too 'bare bones' to be of much use.} \\
         & Not useful & \textit{I like that it only shares public areas, I don't need info on private locations} \\
         & Privacy invasive & \textit{I don’t want my private whereabouts to be tracked to that degree.} \\
         & Acceptable & \textit{I think just general location history would be okay to be released}\\
    \midrule
    Identity & OK for health authorities etc. & \textit{Even though precise identities are given, it's only available to "authorized" people. } \\
        & Not OK for health authorities etc. & \textit{I don't like the information being shared automatically with health workers.} \\
        & OK for tech-savvy users & \textit{it allows more info on diagnosed users for tech savvy people} \\
        & Not OK for tech-savvy users & \textit{App Design 5 seems more useful than the other, and limits the location finding to the public. However, I hate the breach of privacy of App 5 so much (letting tech-savvy people view identities), that I would go with App Design 3.} \\
        & Not OK for anyone & \textit{I'm actually indifferent. Both seem to make specific identities too available to the public.} (when comparing between App 3 and 5.) \\
    \midrule
    Notice method & Prefer app inform exposed users & \textit{App 4 collects less data and I think people would prefer to have the app contact them about exposure.} \\
    & Prefer health workers inform exposed users & \textit{Direct health care contact is important, you can’t rely on people to do quarantine on their own} \\
    \midrule
    Misc. & Less info leads to better adoption & \textit{I think the app with the least information shared with others will be used by more of the population.}\\
        & Toss-up condition & \textit{Both of these are bad choices, but at least App Design 3 doesn't let tech-savvy people view precise identities.} \\
        & No matching explanation & \textit{More inclusive data}\\
    \bottomrule
    \end{tabular}
    \endgroup
    \label{tab:all_code_examples}
\end{table}

The first category ``Location'' is related to comparisons between different levels of location sharing and the corresponding functionality of the app.
Figure \ref{fig:all_code} suggests that both utility and privacy are important factors that affect people's choices.
Most participants considered the functionality of releasing hotspots to be useful, and only a few responses mentioned that collecting more private location information is not useful.
The number of mentions of the code ``Location-privacy invasive'' was 41.8\% more than the code ``Location-acceptable.''

The second and third category ``Identity'' and ``Notice'' are both related to the comparisons between centralized and decentralized designs, corresponding to the implications on privacy and utility respectively. The distinct number of mentions of these two categories suggest that privacy concerns seem to be a more important factor than utility in this design dimension. In the category ``Identity'', both the ``Not OK for health authorities'' and ``Not OK for tech-savvy users'' received a considerable number of mentioning, which suggests neither the centralized design nor the decentralized design address the privacy concerns perfectly.

The last high-level category (``Misc.'') includes three special codes that showed up less frequently or discuss different aspects from other codes, but are still interesting to report. The first one ``Less info leads to better adoption'' shows that some participants chose the app they prefer to install not based on their personal preferences, but on which design better matches the crowd preferences that they assumed. The second one ``Toss-up condition'' shows that sometimes people find two conditions to be equally good or bad. The last code ``No matching explanation'' is used when the responses are too vague to be understood (no information) or seem to convey ideas that seem to contradict the design choices (mismatch). There were 3\% (22 out of 750) mismatched responses in total and some reflect misconceptions about the app designs. This will be further discussed in the limitations.

\subsection{Clustering analysis about individual-level preferences: People who value privacy more than utility tend not to choose any of the six designs}
\label{sec:clustering_analysis}

To understand how individual differences play out in the preferences about contact-tracing apps, we conducted a hierarchical cluster analysis using a bottom-up approach~\cite{rokach2005clustering} on our participants, each represented by a 6-dimensional vector of their absolute willingness-to-install ratings of the six app design options.
To calculate the distance measure, we recoded the 5-point agree level to 3-point, by merging strongly disagree and disagree into one class (-1, ``disagree to install''), and merging strongly agree and agree into one class (1, ``agree to install''). The neutral rating is coded as 0. 

We examined the clustering analysis results with cluster number ($K$) varying from 1 to 10, and determined on $K=5$, as it resulted in a coherent pattern within each cluster and different patterns among these clusters.
In addition, by clustering participants into 5 groups, all clusters have sufficient instances to analyze.
The largest cluster (Cluster-1) contains 66 people (31.7\%);
Cluster-2 contains 52 people (25\%);
Cluster-3 contains 40 people (19.2\%);
and the smallest clusters (Cluster-4 and Cluster-5) each contain 25 people (12.0\%).

\begin{figure}[h]
    \centering
     \includegraphics[width=0.5\linewidth, valign=t]{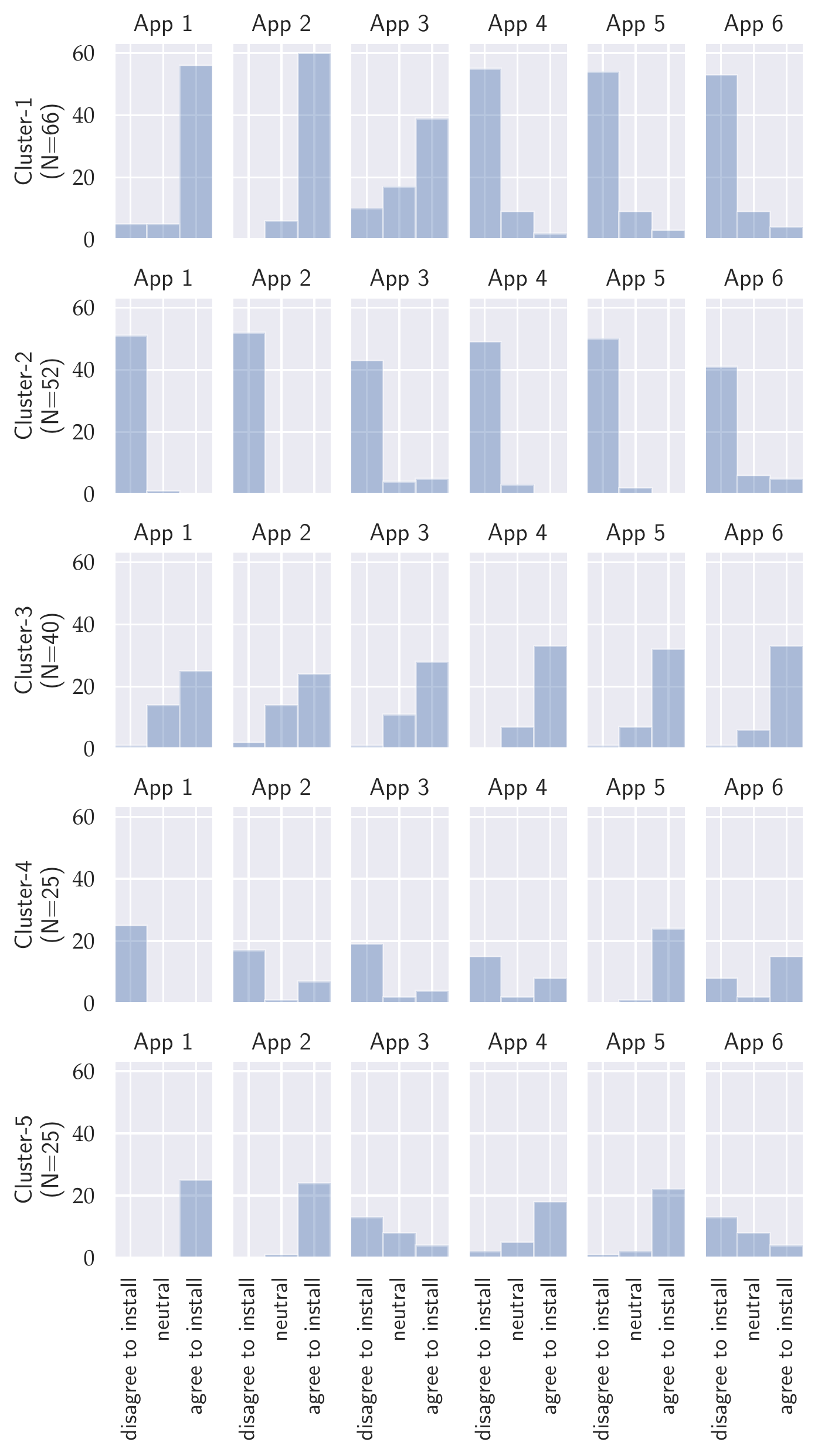}
     \hspace{0.02\linewidth}
     \includegraphics[width=0.45\linewidth, valign=t]{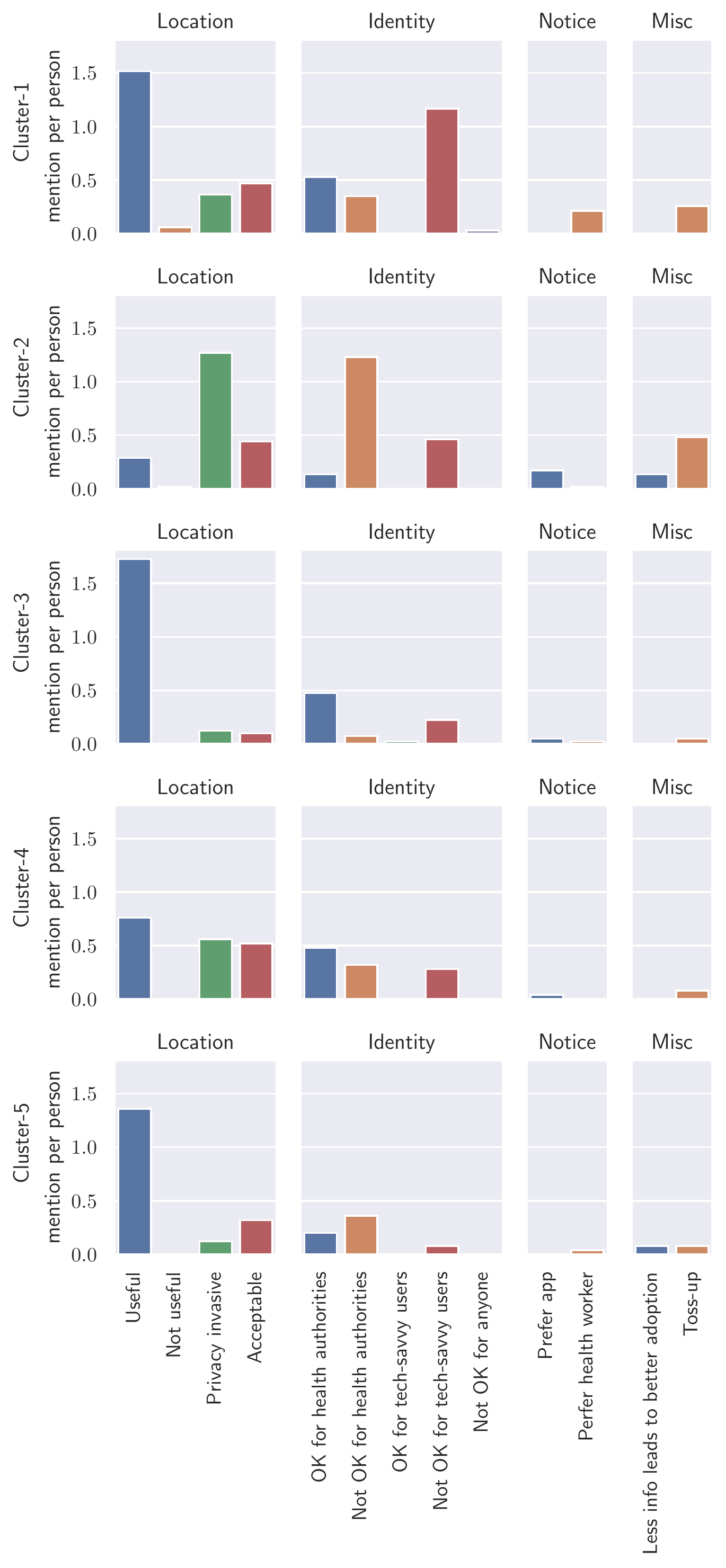}
        \caption{Participants are clustered based on their willingness to install the six types of apps. The clusters are ranked based on their size (N). The left figure shows the aggregated willingness to install the six apps of the five clusters. The right figure shows the average mentions per person of the qualitative codes from the free-form responses (see Section \ref{sec:qualitative_codes}). This analysis helps us identify five groups of people holding different attitudes towards these app designs. For example, Cluster-1 shows that a large group of people strongly prefer centralized designs and strongly dislike decentralized designs; Cluster-2 suggests that people who value privacy more than utility tend not to choose any of the six designs.}
    \label{fig:app_install_cluster}
\end{figure}

Figure \ref{fig:app_install_cluster} presents the clusters ($K=5$) ranked based on their size (from the largest to the smallest). The left figure demonstrates the aggregated willingness to install the six apps for each cluster. The heights of the bar are corresponding to the number of participants. The right figure demonstrates the average mentions per person of the qualitative codes derived from the free-form responses (see Section \ref{sec:qualitative_codes}).

Cluster-2 and Cluster-3 reflect two opposite types of people.
The former dislikes almost all six options, and the latter are neutral or willing to install almost all six options.
The plots of the qualitative codes reveal that people in Cluster-2 had very different feelings about Location and Identity than other clusters, which characterized them as caring more about their privacy than the average person.
First, in the Location category, their responses seldom mentioned location information being useful and often bring up privacy concerns about location collection.
Second, in the Identity category, their responses mentioned substantially more about not feeling comfortable sharing identity information with health workers and health authorities, and relatively less about sharing identity information with tech-savvy users.
In contrast, Cluster-3 participants acknowledged the usefulness of location data more often in their free-form answers, and rarely discussed the privacy concerns regarding sharing location data (``Location-Privacy invasive'') and sharing identity information (``Identity-Not OK for health authorities'' and ``Identity-Not OK for tech-savvy users'').

On the contrary, Cluster-1, 4, 5 present people who have different feelings about the six app designs, which may help inspire better COVID-19 app designs that can potentially reconcile the conflicting interests.

Cluster-1 is the largest cluster obtained from our analysis, which captures a group of people that strongly supported centralized designs and opposed decentralized designs. According to the qualitative coding results, these people had profound concerns about sharing identity information with tech-savvy users, which is a potential risk of decentralized contact-tracing apps. Some of them also preferred to be informed by health workers, which is a functionality supported by centralized contact-tracing apps.

Cluster-4 characterizes people who had a marginal preference in decentralized designs than centralized designs.
Cluster-5 characterizes people who preferred apps that collect and share location of the infected users to the public than apps that do not collect any location. The qualitative coding results showed a similar trend, with the ``Location-Useful'' code mentioned a lot, and the ``Location-Privacy invasive'' code rarely mentioned.

\subsection{Participants showed consistent app installation preferences across the survey.}

We tested the consistency of the relative app installation preferences (measured in pairwise comparison questions) and the absolute app installation preferences (measured in willingness-to-install questions) to gain more understanding into the validity of our results. To achieve this, we derived another set of pairwise comparison results by comparing the absolute willingness-to-install ratings, and then calculated a mismatch score for each person by adding up how many pairs have different results between the original and the inferred pairwise comparison results. If there is a tie, we always consider it to be a match. For example, if a person answered ``strongly agree'' to installing both App 1 and App 2, then no matter which one they chose when answering the pairwise comparison questions, we always considered the relative preferences and absolute preferences to be consistent.

We got a median mismatch score of 1.0, which means that out of the total 15 pairs, more than half of the participants had no more than 1-pair difference between the original and inferred pairwise comparison results. This suggests that our participants showed consistent install preferences across the survey, which justified our choice of combining the absolute willingness-to-install ratings and the free-form responses explaining the relative install preferences to understand people's feelings about these app designs in Section \ref{sec:clustering_analysis}.

\subsection{Limitations}
As in any work, our study has a few caveats that need to be taken into account when interpreting the results.

First of all, our sample size is small and may not be very representative as it was collected on Amazon Mechanical Turk~\cite{kang2014privacy}. We do not have data in some less populated states such as Montana and and Wyoming, and we have more female (56.7\%) than male participants(42.3\%). We did not collect data from people younger than 24 years old, and our sample has higher education level than the general U.S. population (36\% bachelor degree or above for people over age 25 in 2019).

To not overwhelm participants by giving them too much information, our survey simplifies the behavior of contact-tracing apps.
For example, our table says that tech-savvy users can infer nearby infected users' identities without going into details about how they are able to do that.
This could lead participants to estimate higher risks of disclosing data to tech-savvy users than what can actually happen.
Also, throughout the survey, we did not mention about data breaches. They are important security risks, particularly for the centralized case.

Since our main goal is to compare multiple design options, the user may show more variance to reflect their preferences among different designs when rating the willingness to install multiple designs at the same time.
Therefore, the installation rate may not be the same as just asking a single app design option.

Misunderstanding of concepts in the survey could be another cause for inaccuracy in the result.
Our analysis of the free-form explanations of pairwise comparison choices showed that around 3\% (22 out of 750) explanations did not match their selections.

% \jason{Minor note, after uploading to Arxiv, share with the Ireland folks, update the spreadsheet of COVID-related studies if applicable, ping Jodi about the study to see if it's relevant to add to the CMU COVID dashboard, and ping Byron Spice too. And add to your COVID web site, along with all materials.}

\section{Discussion of Survey Findings and Design implications}

\subsection{Centralized contact tracing (with strong security) seems to be a better option at the national level.}
In Section \ref{sec:overall_preference_results}, we showed that App 2, the design that uses centralized contact tracing and release hotspots in public places was the most preferred option to be installed.
There are 55\% of participants that agreed or strongly agreed with the statement: ``If App 2 is available, I would install it on my phone'', which is close to results of the Washington Post and the Ipsos polls (around 50\% smartphone users agree to install)~\cite{Washingt23:online, American37:online}.
Note that this number was achieved when the participants have gone through an in-depth overview of possible privacy risks in these apps, as compared to the other two studies that did not emphasize privacy risks in particular.

Contrary to the assumptions of some prior work, we found that people preferred to install apps using centralized designs to achieve contact tracing.
Our analyses revealed two possible reasons, both related to privacy concerns.
First, both of our analyses on ratings of privacy-related factors (Section \ref{sec:privacy_related_factors}) and qualitative coding analyses (Section \ref{sec:qualitative_codes}) showed that people feel less comfortable disclosing sensitive information to tech-savvy users than to the central authorities. Some people considered central authorities as trustworthy because they are \textit{```authorized'' people'}, some considered it was not ideal to share their identity information with either party, while leaking to tech-savvy people is perceived as a more severe threat.
Second, our clustering analysis (Section \ref{sec:clustering_analysis}) revealed a special type of user (Cluster-2) that did not seem to favor any of the six apps. They may find sharing identity information with the central authorities more concerning and prefer decentralized contact-tracing apps when having to make a choice between decentralized and centralized designs. However, neither of these two options could provide the level of privacy protection that is sufficient for them to use the app.

Our results seem to suggest that centralized contact tracing will be adopted by more people at the national level.
That being said, we also showed that the effect sizes may vary with the partisan leanings of which state the participants currently reside in.
Due to the lack of data in some less populated states (e.g., Wyoming, Montana), we can not fully characterize how install preferences vary in different regions. While we consider more dedicated studies would be helpful, especially if some states finally choose to deploy their own apps (e.g., the app ``Care19'' in North Dakota~\cite{Care19ND41:online}).

Another important takeaway message to health officials, researchers, and journalists is that people may not only have privacy concerns with data flows to central authorities, but also with data flows to the people around them. The latter can be equally scary, if not more, because they could be people the user personally know of, and these people's behaviors will be less under control. Therefore, the public also need to be educated about the risks in decentralized contact-tracing apps, and apps should feature conspicuous privacy notices about these risks, if/when a decentralized contact-tracing app is built and deployed.

Since more sensitive information will be stored on central servers when adopting centralized solutions, there will be even higher requirements for data security. Strong security protection must be applied to prevent data breaches and the access to any user data must be strictly controlled. Whoever develops the app should clearly describe to users what purposes the data will be used for, who may be able to learn the information, and how long it will be stored, and make sure to comply with them across the entire life cycle of the data.

\subsection{Providing more helpful information (e.g., hotspots in public places) may incentivize more people to install the app.}

Although people agreed that the basic contact tracing functionality is helpful (Section \ref{sec:utility_related_factors}), the additional information about hotspots of the disease in public places showed significant effect in increasing installation rate (Section \ref{sec:overall_preference_results}). Many people considered the hotspot feature based on location data helpful and referred to this as part of the reason that they preferred apps that collect location data (Section \ref{sec:qualitative_codes}). As put by one of our participants: \textit{``Again showing location history is going to help more in containing the spread. This sharing of information outweighs the location privacy because it can help show where infected people have been.''} Our results suggest that the value of the hotspot information is straightforward to people, and may make the app more appealing to use.

Another reason is that people generally found only collecting and sharing locations in public areas seem to strike an appealing balance between the privacy risks and utility benefits. Our analyses regarding individual factors showed that collecting locations in public areas is significantly more acceptable to users (Section \ref{sec:privacy_related_factors}), and the perceived usefulness of the information about hotspots in public areas is even significantly higher than hotspots in all places.

Therefore, we suggest that these apps could release aggregated, coarse-grained location history of infected users in a way that minimize the risks of re-identifying these people. The app will only collect locations if the user has explicitly granted the consent, and it should provide options to allow users to only disclose their location traces in public places. This progressive opt-in feature forms new design options that are worth exploring in the next step.

\section{Conclusion}

Our survey provides unique insights into user's opinions on contact-tracing app designs when they have been informed of the privacy risks and utility benefits.
From the results of our survey, we proposed the following suggestions for future privacy-focused COVID-19 contact-tracing apps targeting the U.S. population:
The app should have a proximity-based centralized contact-tracing architecture (with strong security) and leak minimal information about diagnosed users to the public.
The app should also collect user's information on install to prevent malicious users from creating multiple accounts and infer the identity of diagnosed users.
The app may adopt an opt-in location-tracking feature to nudge more people to install the app and allow users to only share their location traces in public areas to reduce privacy risks.
The app should be transparent about both the risks of disclosing the information to governments (for centralized designs) and tech-savvy users (for decentralized designs), and decentralized solutions should not be posited as a risk-free solution.

Hopefully, with these suggestions, digital contact-tracing solutions can help more people in a privacy-friendly way and people can make an informed decision on installing those apps.

\bibliographystyle{unsrt}  
\bibliography{other,hcct}  %%% Remove comment to use the external .bib file (using bibtex).
%%% and comment out the ``thebibliography'' section.

% \appendix

\end{document}